\documentclass[preprint,12pt]{elsarticle}

\usepackage{fullpage}
\usepackage[colorlinks=true]{hyperref}
\usepackage{listings}
\usepackage{float}

\usepackage{amssymb, amsmath}
\usepackage{amsthm}

\journal{Journal of Computational Physics}

\usepackage{tikz}
\usetikzlibrary{shapes,arrows}
\usepackage[normalem]{ulem}
\usepackage{hhline}

\usepackage{graphicx}
\usepackage{subfig}
\usepackage{color}

\usepackage{pgfplots}
\usepackage{pgfplotstable}
\definecolor{markercolor}{RGB}{124.9, 255, 160.65}
\pgfplotsset{
compat=1.3,
width=10cm,
tick label style={font=\small},
label style={font=\small},
legend style={font=\small}
}

\usetikzlibrary{calc}
\usetikzlibrary{intersections}

\begin{document}

\begin{frontmatter}



\title{Fractional models of Reynolds-averaged Navier-Stokes equations for Turbulent flows}


\author[Brown]{Pavan Pranjivan Mehta}\corref{cor1}

\cortext[cor1] {Corresponding author.~E-mail address: pavan\_pranjivan\_mehta@brown.edu}

\address[Brown]{Brown University, 170 hope St., Providence, Rhode Island, United States of America, 02906}

\begin{abstract}

Its is a well known fact that Turbulence exhibits non-locality, however, modeling has largely received local treatment following the work of Prandl over mixing-length model. Thus, in this article we report our findings by formulating a non-local closure model for Reynolds-averaged Navier-Stokes (RANS) equation using Fractional Calculus. Two model formulations are studied, namely one- and two-sided for Channel, Pipe and Couette flow, where the results shown have less 1\% error. The motivation of two-sided model lies in recognising the fact that non-locality at a given spatial location is an aggregate of all directions. Furthermore, scaling laws and asymptotic relationship for Couette, Channel and Pipe flow is reported. It is to be noted that modeling in wall units, no additional coefficient appears, thus there models could be applied to complex flows with ease.

\end{abstract}


\end{frontmatter}



\section{Introduction} \label{intro}
%
%

Direct Numerical Simulations (DNS) aim's at resolving all scales but are limited to simple flow configuration as the grid requirements are $\approx Re^{9/4}$ \cite{pope_2000}, thus for complex flows it exceeds today’s computational ability. Large Eddy Simulations (LES) addresses this problem by resolving only the high energy modes by spatially filtering the fields, while modeling the smaller scales. The pioneering work of Osborne Reynolds of time-averaging the Navier-Stokes equations, known as Reynolds-averaged Navier-Stokes (RANS) equations; further brought down the computations to acceptable limits \cite{pope_2000} and has been an industrial work-horse ever-since. As a result of not resolving all the scales, a closure problem arises both in the case of LES and RANS. Despite the evidence of non-locality in Turbulence, modeling has largely received local treatment motivated by the early work of Prandl over the mixing length model. Plagued with modeling assumptions for the unclosed terms, they are susceptible to failure and are not generalisable, as evident by a number of studies such as \cite{craft1993impinging, wilcox1988reassessment}. Thus, in this study, we propose to employ Fractional calculus for non-local modeling of Channel, Pipe and Couette flow.

Fractional calculus is regarded as a generalisation of the integer-order calculus. The first ever recorded in history with curiosity of L'hospital for differentiation with the order, $n = 1/2$, following Lebniz notation ($d^n / dy^n$) for $n^{th}$ order differentiation. Replying to his letters in 1695, Leibniz mentioned it as an apparent paradox for which one day will have useful consequences. A more detailed history is given in chapter 1 of \cite{Ross1975}.

To the author's knowledge the earliest text known addressing non-locality in fluids was by Richardson \cite{richardson1926atmospheric}, which inspired the early work over the subject. He used the distance - neighbor graph to discover a power-law behaviour within the atmospheric diffusion process where a non-Fickian equation was constructed as opposed to Fick's law of diffusion with a constant diffusivity. The work of Richardson inspired three directions which today are the corner stone for non-local modelling of Turbulence. The earliest attempt with regards to modelling was recorded by Kranichnan \cite{kraichnan1987eddy}, however it did not use Fractional derivatives. It was Chen \cite{chen2006speculative} who first proposed the use of Fractional derivative of a constant order with connections to Richardson's work \cite{richardson1926atmospheric}. However, a revision is required over the numerical value from Richrdson's work with modern instrumentation, data and techniques. Further remark's over these two works will be detailed in preceding sections. Chen's work was furthered in \cite{song2018universal} with a proposition of a variable-order fractional closure model in non-divergence form. The limitation of the previous study was mitigated by formulating a novel model in divergence form \cite{mehta2019discovering}, where it was showed that no-coefficient is required when non-dimensionalised in wall units. Another interesting direction is within the framework of Lattice Boltzmann Method (LBM) laid in \cite{epps2018turbulence, hayot1996non}. Besides Chen \cite{chen2006speculative}, these directions are of fundamental importance for the development of Turbulence. It is to be noted that although the proposition may be of a model, however, it should not be regarded merely a modeling effort but rather a fundamental theory of Turbulence and modeling is merely an application of their conjectures as we shall discuss throughout this paper. 

This discussion is not complete without mentioning Hamba's work \cite{hamba1995analysis, hamba2004nonlocal, hamba2005nonlocal, hamba2006mechanism, hamba2017history} following Kranichnan \cite{kraichnan1987eddy}, which details numerical results. The key inference here is that there is no unique non-local kernel even for turbulent flows, but rather depended on the Spatial location. This indicates the presence of multiple-scales and proves that the non-locality is not a constant in turbulent flows as initially proposed by Chen \cite{chen2006speculative} where he proposed the use of fractional derivative of a fixed order. The current work indeed proves the same fact, where the use of a variable-order fractional closure model implies that the non-locality is spatially dependent. It is to be noted that, there is a limitation to Kranichnan's and Hamba's work, where the non-local operator constructed cannot approximate the local or laminar regions, thus this model may not be suitable for modeling wall-bounded flows, where the near wall regions are dominated by the viscous actions. However, as we show our model is valid for all regimes of the flow.

\section{Fractional Reynolds-averaged Navier-Stokes equations (f-RANS)} \label{data_sets}
%
%



In this section, we introduce the fractional closure model for fluid flows for cases where statistical stationarity is achieved, needless to say they are valid for unsteady flows too as the non-locality is considered in space rather than time. The governing equations are steady state Reynolds-averaged Navier-Stokes (RANS) equations (\ref{eq:rans}) \cite{mehta2019discovering}, where $\overline{(.)}$ implies that quantities are temporal-averaged. The instantaneous velocity ($U_i$) and pressure ($P$) is given by $U_i = \overline{U_i} + u_i$ and $P = \overline{P} + p$, where $u_i$ and $p$ are the fluctuating components of velocity and pressure about the mean. The time-period, $T$ for the ensemble is large enough to satisfy, $ {1 \over  T} \int_0^T u_i = 0$.

\begin{equation}\label{eq:rans} 
\overline{U^{+}_{j}} \frac{\partial \overline{U^{+}_{i}}} {\partial x^{+}_{j}} = -\frac{\partial \overline{P^{+}}} {\partial x^{+}_{i}} + \frac{\partial} {\partial x^{+}_{j}} \left( \frac{\partial \overline{U^{+}_{i}}} {\partial x^{+}_{j}} - \overline{u_{i}u_{j}}^{+} \right) ~;~ i,j = 1,2,3
\end{equation}

The above RANS equations (\ref{eq:rans}) is written in non-dimensional form (in wall units) using the friction velocity ($U^2_\tau = \tau_w / \rho$) and kinematic viscosity ($\nu$), where $\tau_w$ is the wall shear stress. The notation $(.)^+$  indicates the same and as given in (\ref{eq:non-dim-new-model}). 

\begin{equation}\label{eq:non-dim-new-model}
{\overline{U^{+}_{i}}} = \frac{ \overline{U_{i}}} {U_{\tau}} ~;~  {x^{+}_{i}} = \frac{x_{i} U_{\tau}}{ \nu} ~;~ (\overline{u_{i}u_{j}})^{+} = \frac{\overline{u_{i}u_{j}}}{ U^{2}_{\tau}} ~;~ {\overline{ P^{+}}} = \frac{ \overline{P}} {\rho U^{2}_{\tau}}~;~ {\tau^{+}} = \frac{\tau} {\rho U^{2}_{\tau}} .
\end{equation}


The generalised fractional calculus model for the unclosed terms in (\ref{eq:rans}) takes the general form of (\ref{eq:frans}) and shall be referred to as fractional Reynolds-averaged Navier-Stokes equations (f-RANS). The fractional order ($\alpha$) is spatially dependent function in $x^+_j$, as turbulence exhibits multiple scales, where $\alpha$ is co-related to turbulence length-scale. In the absence of the Reynolds stresses ($\overline{u_{i}u_{j}}$), it is trivial to show that the fractional order is unity. This is an important remark for both wall-bounded and transitional regimes of the flow, as there are significant regions where the flow is dominated by the viscous actions of the fluid. Thus this model is valid for all regimes of the flow. Furthermore, any additional term in (\ref{eq:rans}) to account for unsteadiness in the flow or body forces would remain "as is" in (\ref{eq:frans}) and our proposition in  (\ref{eq:frac}) would remain valid as long as (\ref{eq:rans}) remains valid, although we do not prove it here and leave it as future work. However, we justify it by recognising the fact that the non-locality resides in the Reynolds stresses ($\overline{u_{i}u_{j}}$), which may not be unique but (\ref{eq:rans}) remains valid, regardless. Thus, it is expected that the fractional order ($\alpha$) is not unique for different flows, however, (\ref{eq:frac}) remains valid.

\begin{equation}\label{eq:frac} 
    ^M D_{x^+_j}^{\alpha(x^+_j)} (\overline{U^+_i})  := \frac{\partial \overline{U^{+}_{i}}} {\partial x^{+}_{j}} - \overline{u_{i}u_{j}}^{+}~;~ i,j = 1,2,3 ~;~ \alpha(x^+_j) \in (0, 1]
\end{equation}

\begin{equation}\label{eq:frans} 
\overline{U^{+}_{j}} \frac{\partial \overline{U^{+}_{i}}} {\partial x^{+}_{j}} = -\frac{\partial \overline{P^{+}}} {\partial x^{+}_{i}} + \frac{\partial} {\partial x^{+}_{j}} \left(  ^M D_{x^+_j}^{\alpha(x^+_j)} (\overline{U^+_i}) \right) ~;~ i,j = 1,2,3 ~;~ \alpha(x^+_j) \in (0, 1]
\end{equation}

We further formulate (\ref{eq:frac}) as one-sided model (\ref{eq:one}) defined over the domain $[-\infty, x^+_j]$ and two-sided model (\ref{eq:two}) defined over the domain $[-\infty, \infty]$, where for notational convenience, we use (\ref{eq:frac}) to imply inferences applicable to both one- and two-sided model. The choice of either model will be justified in the forthcoming sections of the this article, where we apply it to Channel, Couette, Pipe flows.

\begin{equation}\label{eq:one} 
    {_{-\infty}} ^C D_{x^+_j}^{\alpha(x^+_j)} ( \overline{U^+_i}) := \frac{\partial \overline{U^{+}_{i}}} {\partial x^{+}_{j}} - \overline{u_{i}u_{j}}^{+}~;~ i,j = 1,2,3 ~;~ \alpha(x^+_j) \in (0, 1]
\end{equation}

\begin{align} \label{eq:two}
\begin{split}
    {_{[-\infty, \infty]}} ^T D_{x^+_j}^{\alpha(x^+_j)} ( \overline{U^+_i}) ~~&:=  {1 \over 2 }~ ({_{-\infty}} ^C D_{x^+_j}^{\alpha(x^+_j)} ~-~~  {_{x^+_j}} ^C D_{\infty}^{\alpha(x^+_j)} ) ~\overline{U^+_i}  \\ ~~&:= \frac{\partial \overline{U^{+}_{i}}} {\partial x^{+}_{j}} - \overline{u_{i}u_{j}}^{+} ~;~ i,j = 1,2,3 ~;~ \alpha(x^+_j) \in (0, 1]
\end{split}
\end{align}

\noindent
${_{-\infty}} ^C D_{x^+_j}$ and ${_{x^+_j}} ^C D_{\infty}^{\alpha(x^+_j)}$ is the left- (\ref{eq:left}) and right- sided (\ref{eq:right}) Caputo fractional derivative respectively, where $\alpha(x^+_j) \in (0, 1] $.

\begin{equation}\label{eq:left} 
   {_{-\infty}} ^C D_{x^+_j}  \overline{U^+_i} = \frac{1}{\Gamma(1-\alpha(x^+_j))}\int_{-\infty}^{x^+_j} (x^+_j-\xi)^{-\alpha(x^+_j)}\frac{ d \overline{U^+_i}(\xi)}{d\xi}d\xi
\end{equation}

\begin{equation}\label{eq:right} 
   {_{x^+_j}} ^C D_{\infty}^{\alpha(x^+_j)} \overline{U^+_i} = \frac{1}{\Gamma(1-\alpha(x^+_j))}\int_{x^+_j}^{\infty} (\xi - x^+_j)^{-\alpha(x^+_j)}\frac{d\overline{U^+_i}(\xi)}{d\xi}d\xi
\end{equation}

 It is to be noted that we model the total shear stress as opposed to the common practice of only modeling the Reynolds stresses. It is evident whilst modeling in wall units, it is free from all assumptions, thus this model is generlisable to any flow. Since, there appears no coefficient to consider, this model can be readily applied to any complex three-dimensional flows. 

In the forthcoming sections, we shall outline the application of f-RANS model to Couette, Channel and Pipe flow.

\subsection{Turbulent Couette Flow}

Couette Flow is characterised by flow between two infinitely long parallel plates \cite{avsarkisov2014turbulent}, where one of them is moving with a constant velocity in the streamwise direction ($x^+$), while the other is at rest with zero pressure gradient. This implies that there is only a finite streamwise component of velocity ($U^+$) and wall-normal ($y^+$) component of velocity ($V^+$) is infinitesimally small. The spanwise direction ($z^+$) is treated as homogeneous with little or no-variation in the flow quantities, thus mean spanwise velocity component ($W^+$) and all gradients in spanwise direction vanish. After simplifying the RANS equations (\ref{eq:rans}), the governing equation for the turbulent Couette flow is given as follows (\ref{eq2.3}),

\begin{align}\label{eq2.3} 
\frac{d} {d y^{+}} \left( \frac{d \overline{U^{+}}} {d y^{+}} - (\overline{uv})^{+} \right) = 0.
\end{align}

\noindent
Here, $U^+$ is a function of $y^+$ due to $\frac{\partial U^+}{\partial x^+}=0$. Upon integrating (\ref{eq2.3}) with respect to $y^+$, produces (\ref{eq2.4}),

\begin{equation}\label{eq2.4}
  \frac{d \overline{U^{+}}} {d y^{+}} - (\overline{uv})^{+} = C.
\end{equation}

The constant, $C$ is evaluated at the wall (within the viscous sub-layer), where the Reynolds's stresses are negligible, namely, $(\overline{uv})^+|_{y^+=0} \approx 0$, and $\overline{U^+} \approx y^+$. Thus, we have $C=(d \overline{U^{+}}/{d y^+})|_{y^+=0} = \tau_w^+=1$. Finally, the simplified RANS equation for the turbulent Couette flow is written as (\ref{eq2.5}),

\begin{align} \label{eq2.5}
   \frac{d \overline{U^{+}}} {d y^{+}} - (\overline{uv})^{+} = 1.
\end{align}

The f-RANS one-sided model is defined over the domain $[0, y^+]$, where $ y^+ \in (0 , Re_{\tau}]$ is given as (\ref{eq2.6}), while the two-sided model (\ref{eq2.7}) is defined over the domain $[0, 2Re_\tau]$, where $ y^+ \in (0 ,2Re_{\tau})$ for turbulent Couette flow. For numerical computation of the fractional order ($\alpha(y^+)$) we employ the DNS database generated in \cite{avsarkisov2014turbulent}.

\begin{equation}  \label{eq2.6}
 ^C_{0}D_{y^+}^{\alpha(y^+)} \overline{U^+} ~:=~ 1 ~;~ \alpha(y^+) \in (0, 1]
\end{equation}

\begin{equation}  \label{eq2.7}
 {_{[0, ~2Re_\tau]}} ^T D_{y^+}^{\alpha(y^+)} \overline{U^+} ~:=~ 1 ~;~ \alpha(y^+) \in (0, 1]
\end{equation}

\subsection{Turbulent Channel Flow}

Channel flow is also flow between two infinity long parallel plates, unlike the Couette flow, both the plates remain at rest and flow is driven by streamwise pressure gradient \cite{lee_moser_2015}. All the other remarks remain similar as the Couette for simplifying the RANS equation (\ref{eq:rans}), thus the governing equation for the turbulent Channel flow is given by (\ref{eq2.3:chal}),

\begin{align}\label{eq2.3:chal} 
\frac{d} {d y^{+}} \left( \frac{d \overline{U^{+}}} {d y^{+}} - (\overline{uv})^{+} \right) = -\frac{\partial \overline{P^{+}}} {\partial x^{+}}.
\end{align}
Here, $\overline{U^+}$ is a function of $y^+$ due to $\frac{\partial \overline{U^+}}{\partial x^+}=0$. Further integrating (\ref{eq2.3:chal}) with respect to $y^+$, produces (\ref{eq2.4:chal})

\begin{equation}\label{eq2.4:chal} 
  \frac{d \overline{U^{+}}} {d y^{+}} - (\overline{uv})^{+} = -\frac{\partial \overline{P^{+}}} {\partial x^{+}} y^+ + C
\end{equation}

The constant $C$ is evaluated at the wall (within the viscous sub-layer), where the Reynolds's stresses are negligible, namely, $(\overline{uv})^+|_{y^+=0} \approx 0$, and $\overline{U^+} \approx y^+$. Thus, we have $C=(d \overline{U^{+}}/{d y^+})|_{y^+=0} = \tau_w^+=1$.
Furthermore, at the center line ($y^+ = Re_\tau$), as the flow is turbulent $(d \overline{U^{+}}/{d y^+})|_{y^+=Re_\tau} = 0$, also $\overline{uv} = 0$ as a consequence of symmetry, this produces $\frac{\partial \overline{P^{+}}} {\partial x^{+}} = {1 / Re_\tau} $, thus the RANS equation for turbulent Channel flow is given by (\ref{eq2.5:chal})
\begin{align} \label{eq2.5:chal}
   \frac{d \overline{U^{+}}} {d y^{+}} - (\overline{uv})^{+} = -{y^+ \over Re_\tau} + 1
\end{align}

The f-RANS one-sided model is defined over the domain $[0, y^+]$, where $ y^+ \in (0 ,Re_{\tau}]$ is given as (\ref{eq2.6:chal}), while the two-sided model (\ref{eq2.7:chal}) is defined over the domain $[0, 2Re_\tau]$, where $ y^+ \in (0 ,2Re_{\tau})$ for turbulent Channel flow. For numerical computation of the fractional order ($\alpha(y^+)$) we employ the DNS database generated in \cite{lee_moser_2015}

\begin{equation}  \label{eq2.6:chal}
 ^C_{0}D_{y^+}^{\alpha(y^+)} \overline{U^+} ~:=~  -{y^+ \over Re_\tau} + 1 ~;~ \alpha(y^+) \in (0, 1]
\end{equation}

\begin{equation}  \label{eq2.7:chal}
 {_{[0, ~2Re_\tau]}} ^T D_{y^+}^{\alpha(y^+)} \overline{U^+} ~:=~  -{y^+ \over Re_\tau} + 1 ~;~ \alpha(y^+) \in (0, 1]
\end{equation}

\subsection{Turbulent Pipe Flow}

The arguments for the Pipe flow \cite{wu2008direct} follows that of the Channel, however we transform the equations in cylindrical co-ordinates \cite{batchelor_2000}. Since, the azimuthal direction for this case is regarded as the homogeneous direction besides the streamiwise direction. Thus, all dependence for the pipe follows a radial direction. The streamwise pressure gradient is also non-zero in this case. Thus the simplified RANS equation for turbulent Pipe flow is given in (\ref{eq2.3:pipe}),

\begin{align}\label{eq2.3:pipe} 
\frac{d} {d r^{+}} \left( \frac{d \overline{U_z^{+}}} {d r^{+}} - (\overline{u_r u_z})^{+} \right) = \frac{d \overline{P^{+}}} {d z^{+}}.
\end{align}

\noindent
Here, $\overline{U_z^{+}}$ is a function of the radial direction, $r^+$. Upon Integrating (\ref{eq2.3:pipe})  with respect to $r^+$ produces (\ref{eq2.4:pipe}),

\begin{align}\label{eq2.4:pipe} 
  \frac{d \overline{U_z^{+}}} {d r^{+}} - (\overline{u_r u_z})^{+}  = \frac{d \overline{P^{+}}} {d z^{+}}. r^+ + C.
\end{align}

At the centre of the pipe ($r^+ = 0$), we have $ d\overline{U_z^{+}}/ {d r^{+}} \approx 0 $ as the flow is dominated by turbulence. As a result of symmetry we have, $\overline{u_r u_z}^{+} = 0$, this results in $C = 0$. Furthermore, near the wall, we have $ d\overline{U_z^{+}}/ {d r^{+}} = \tau_w  =1$, since the flow is dominated by the viscous interactions, we have $\overline{u_r u_z}^{+} \approx 0$, thus (\ref{eq2.5:pipe}), where $R^+$ is the Karman number, which plays the similar role as the friction Reynolds number ($Re_\tau $)

\begin{align}\label{eq2.5:pipe} 
  \frac{d \overline{P^{+}}} {d z^{+}} = {1 \over R^+}
\end{align}

Furthermore, we express the RANS equation for turbulent pipe flow from the wall to the center of the pipe by $ r^+ = ((1- r)/R)^+$, where $r \in [0, R]$, with the radius of the pipe, $R = 1$ thus we have (\ref{eq2.6:pipe}). Upon examining, the right hand-side the similarity with Channel flow is obvious.


\begin{align}\label{eq2.6:pipe} 
  \frac{d \overline{U_z^{+}}} {d ( 1- r)^+} - \overline{u^+_{(1-r)^+}  u^+_z}  = {(1 - r)^+ \over R^+} 
\end{align}

The f-RANS one-sided model is defined over the domain $[0, (1-r)^+]$, where $ (1-r)^+ \in (0 ,R^+]$ is given as (\ref{eq2.6:pipe}), while the two-sided model (\ref{eq2.7:pipe}) is defined over the domain $[0, 2R^+]$, where $ (1-r)^+ \in (0 ,2R^+)$ for turbulent Pipe flow. For numerical computation of the fractional order ($\alpha((1-r)^+)$) we employ the DNS database generated in \cite{el2013direct}

\begin{equation}  \label{eq2.6:pipe}
 ^C_{0}D_{(1-r)^+}^{\alpha((1-r)^+)} \overline{U^+} ~:=~  {(1 - r)^+ \over R^+}  ~;~ \alpha((1-r)^+) \in (0, 1]
\end{equation}

\begin{equation}  \label{eq2.7:pipe}
 {_{[0, ~2R^+]}} ^T D_{(1-r)^+}^{\alpha((1-r)^+)} \overline{U^+} ~:=~  {(1 - r)^+ \over R^+}  ~;~ \alpha((1-r)^+) \in (0, 1]
\end{equation}


\section{Physics-informed Neural Network}

In this section we outline the numerical implementation using Physics-informed Neural Networks (PINNs). PINNs are deep neural network where the loss comprises of two parts, namely, data and equations along with weights to balance the two components of the loss \cite{raissi2019physics, fpinns}. As shown in the expression below (\ref{eq:loss}), the weight, $\lambda$ is a hyperparameter to accelerate the convergence: 
\begin{equation} \label{eq:loss}
    L = \lambda L_d + L_e
\end{equation}
The term $L_d$ is the data part, where we supply the boundary and initial conditions, while the term $L_e$ is for the governing equations \cite{raissi2019physics}, where in this case we employ fractional models. Furthermore, we employ adaptive activation functions to further accelerate convergence substantially and avoid bad minima~\cite{Ameya_AAF}.

\subsection{Inverse Problem}

For the inverse problem the goal is to find the fractional order ($\alpha(y^+)$) given the velocity profile and total shear stress from DNS database or otherwise as outlined in the preceding section for each case. Here we adopt a pointwise strategy, this implies that we train the neural network for a single training point at a time. Naturally, the neural network is much smaller consisting of at most three to four layers and two to three neurons per hidden layer. 

\begin{figure}
\centering
\includegraphics[width=0.6\textwidth]{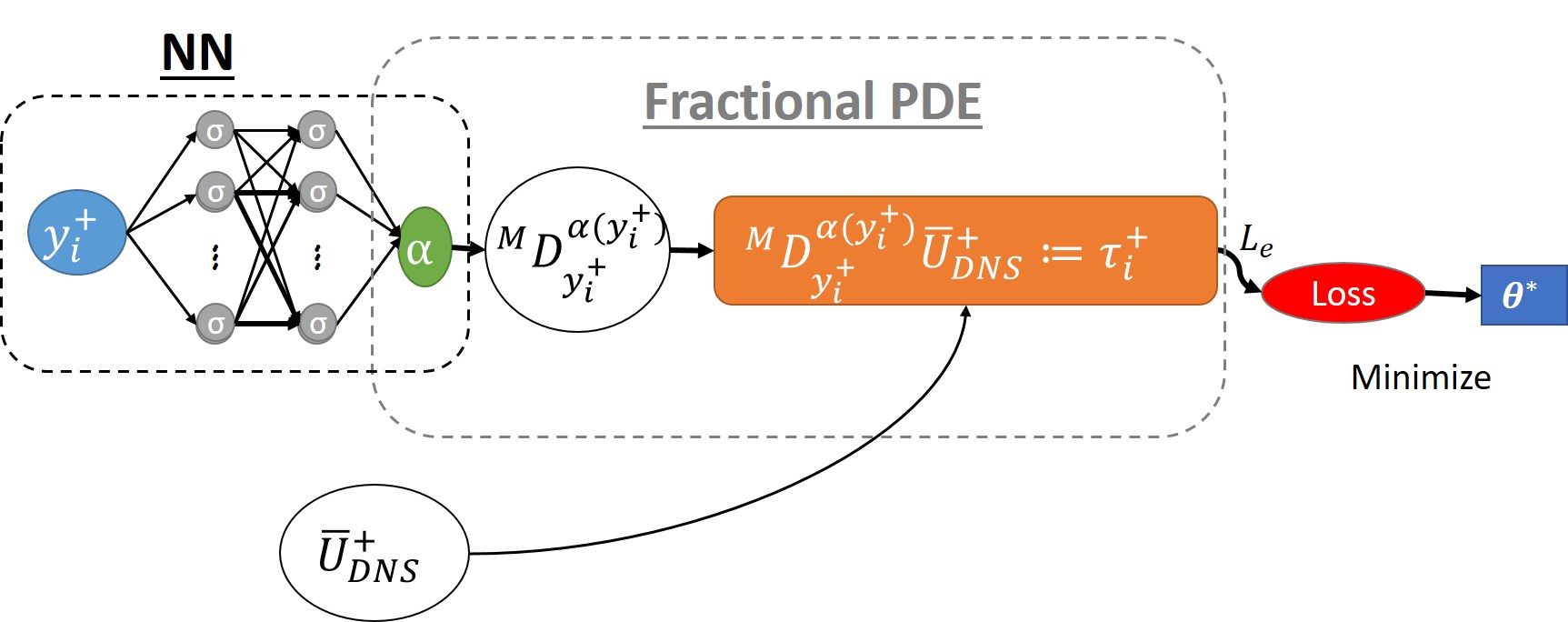}%
\caption{A schematic representation for the inverse modeling, here the spatial location ($y^+$) is the input of the feed forward neural network, while the fractional order ($\alpha$) is the output, which is used to computed the f-RANS model along with velocity from DNS databases}
\label{fig:fpinns}
\end{figure}

Here the loss (\ref{eq:loss}) only comprises of the equation term ($L_e$) given in (\ref{eq:loss_e_al}) written for the $i^{th}$ training point.  A schematic diagram is show in fig.~\ref{fig:fpinns}. A hyperparameter ($\lambda$) may be added to normalize too high or low numbers, however, it does not affect the training process, since it depends on the gradient of the loss rather than its absolute value.  

\begin{equation} \label{eq:loss_e_al}
    L_e = \left[ ^M D_{y^+_i}^{\alpha_{NN}(y^+_i)} (\overline{U^+_{DNS}}) - \tau^+(y^+_i) \right ]^2 
\end{equation}

For numerical discretisation  of the Fractional derivative we employ an {\it L1} scheme \cite{yang2010numerical} for the left- (\ref{eq:left_unn}) and right-sided (\ref{eq:right_unn}) derivative. A uniform finite difference grid was generated by dividing the domain in $M$ points such that the step size, $h$ was kept constant for all the training points, where, $l = 1, 2, ..., M-1$. It is to be noted that the domain for one-sided model is $[0, y^+], y^+ \in (0, Re_\tau]$, while the domain for the two-sided model is $[0, 2Re_\tau], y^+ \in (0, 2Re_\tau)$. Furthermore, we exploit the fact that the Riemann–Liouville and Caputo fractional derivative varies in one term for the fractional order, $\alpha \in (0,1]$.

\begin{align} \label{eq:left_unn}
\begin{split}
    &^C_0 D_{y^+}^{\alpha_{NN}(y^+_i)} (\overline{U^+_{DNS}}) \\ &\approx -{h^{-\alpha_{NN}(y^+_i)} \over \Gamma(2 - {\alpha_{NN}(y^+_i)})} \sum_{j=0}^{l-1} (\overline{U^+}_{{DNS}_{l-j}} -  \overline{U^+}_{{DNS}_{l-j-1}}) [ (j+1)^{1-{\alpha_{NN}(y^+_i)}} -  j^{1-{\alpha_{NN}(y^+_i)}} ]
\end{split}
\end{align}

\begin{align} \label{eq:right_unn}
\begin{split}
    &^C_{y^+} D_{2Re_\tau}^{{\alpha_{NN}(y^+_i)}} (\overline{U^+_{DNS}}) \\ &\approx -{h^{{-\alpha_{NN}(y^+_i)}} \over \Gamma(2 - {\alpha_{NN}(y^+_i)})} \sum_{j=0}^{M-l-1} (\overline{U^+}_{{DNS}_{l+j}} -  \overline{U^+}_{{DNS}_{l+j+1}}) [ (j+1)^{1-{\alpha_{NN}(y^+_i)}} -  j^{1-{\alpha_{NN}(y^+_i)}} ]
\end{split}
\end{align}

\noindent
The merit of pointwsie strategy are as follows:

\begin{itemize}
    \item Given the fractional order exisits, it does not require boundary conditions, which we earlier imposed as unity at the wall \cite{mehta2019discovering}. Thus for more complex flows where the near wall region is not viscous dominated, we can find a physically consistent fractional order.
    \item There are cases, where the fractional order is a discontinuous function or the neural network collapses during training process; a pointwise algorithm mitigates all the limitations. Since, the mean velocity is a smooth and continuous function, we rarely encounter any problems during its training when solved for velocity fields as demonstrated in \cite{mehta2019discovering}.
\end{itemize}

\section{Results: Turbulent Couette, Channel and Pipe Flow} \label{one_two}
%
%

In this section we describe our findings for both one- and two-sided model for turbulent Couette, Channel and Pipe Flow

\subsection{One-sided v/s two-sided formulation}

\begin{figure}[H]
\centering
    \subfloat[one-sided]{{\includegraphics[trim=0.3cm 0.3cm 0.3cm 0.3cm,clip, width=0.5\textwidth]{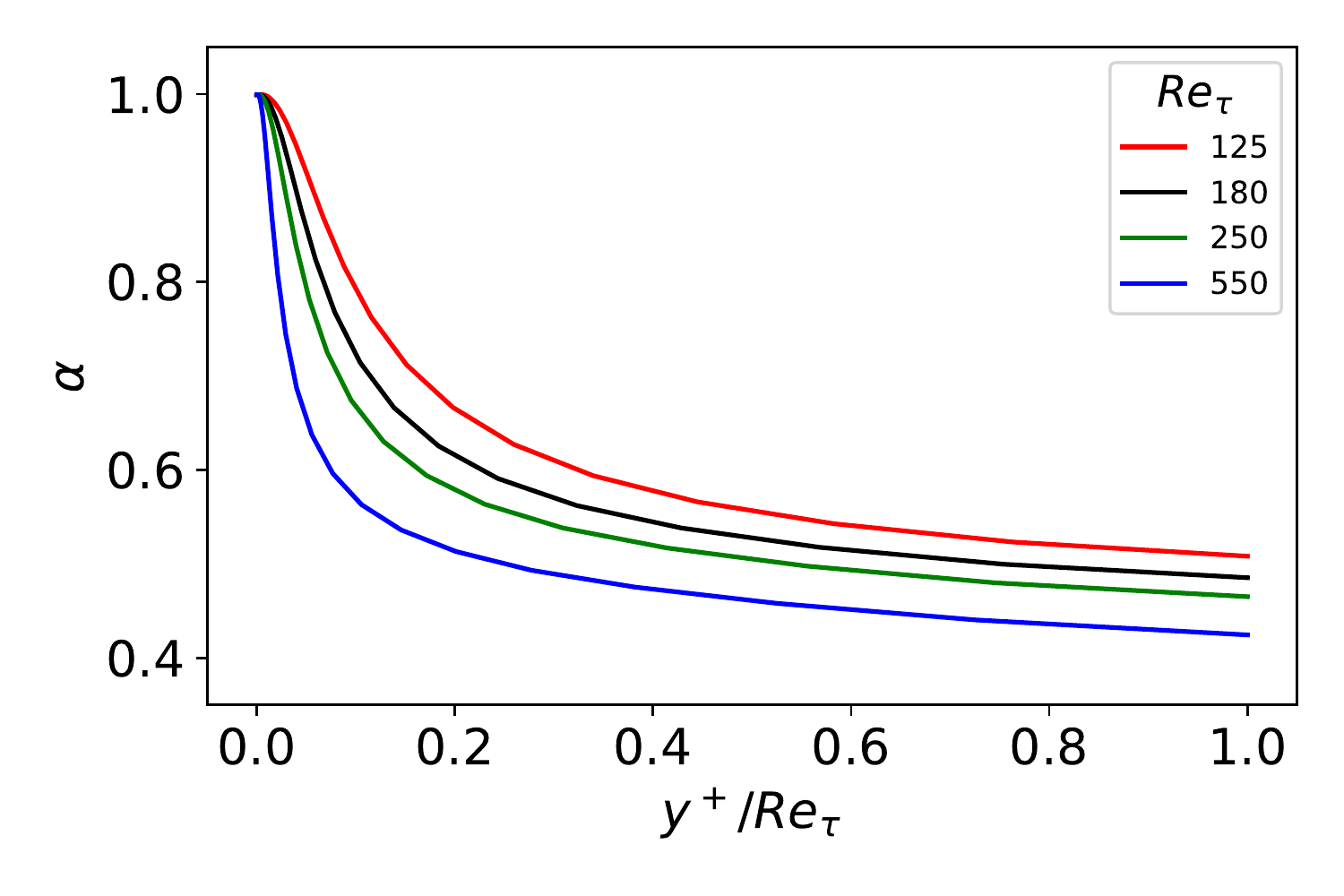}}}%
   \subfloat[two-sided]{{\includegraphics[trim=0.3cm 0.3cm 0.3cm 0.3cm,clip,    width=0.5\textwidth]{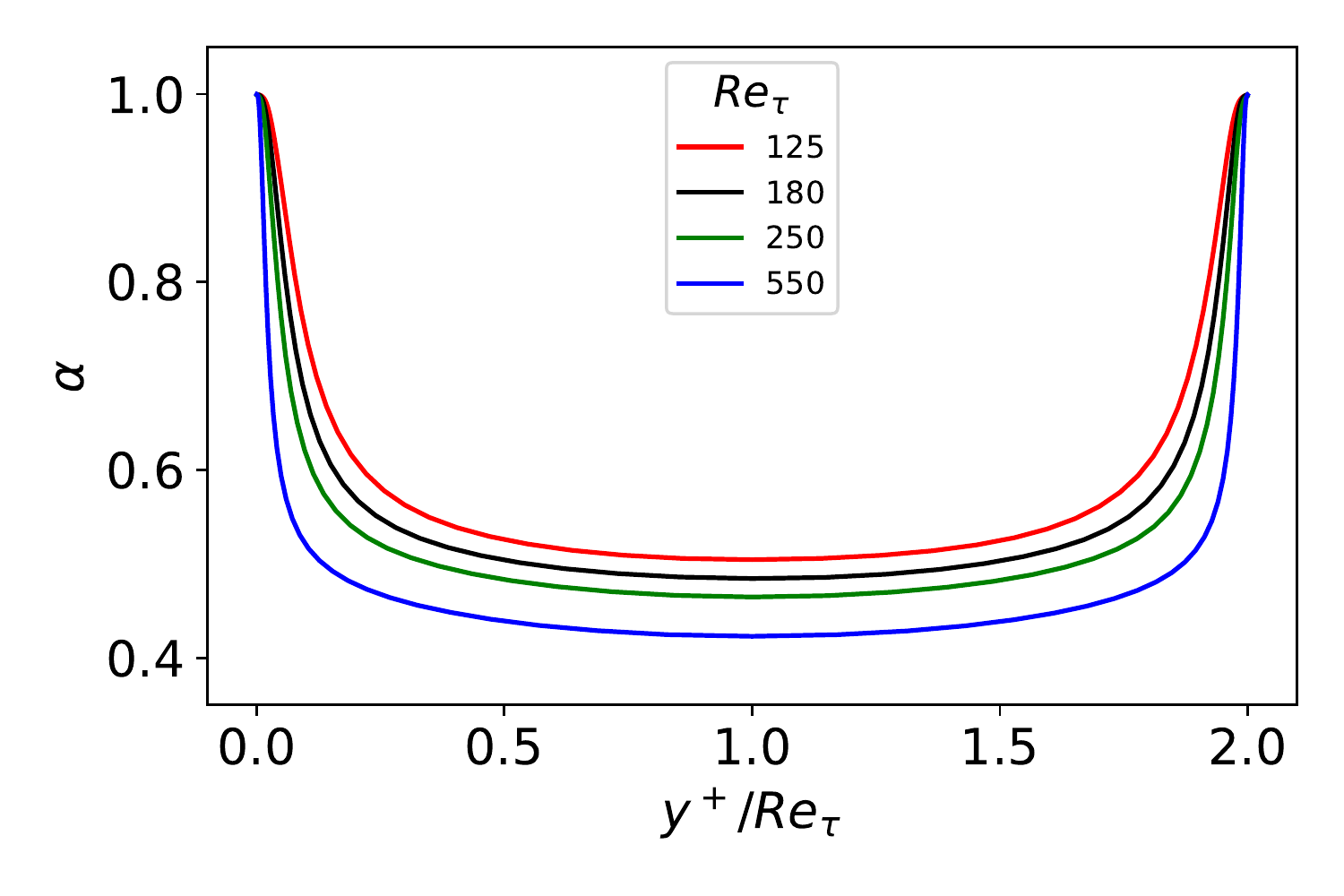}}} \\
   \subfloat[one-sided]{{\includegraphics[trim=0.3cm 0.3cm 0.3cm 0.3cm,clip, width=0.5\textwidth]{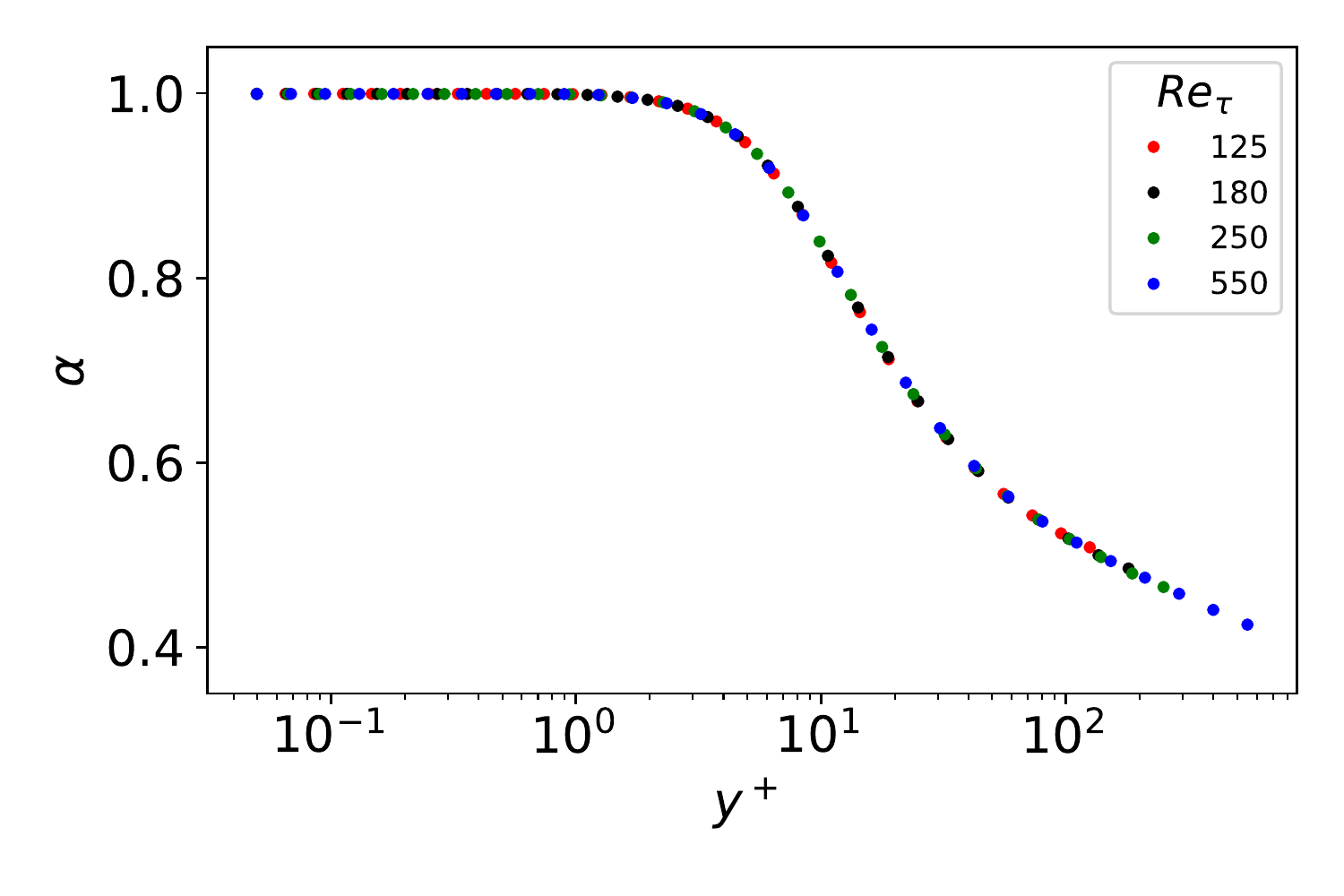}}}%
   \subfloat[two-sided]{{\includegraphics[trim=0.3cm 0.3cm 0.3cm 0.3cm,clip,    width=0.5\textwidth]{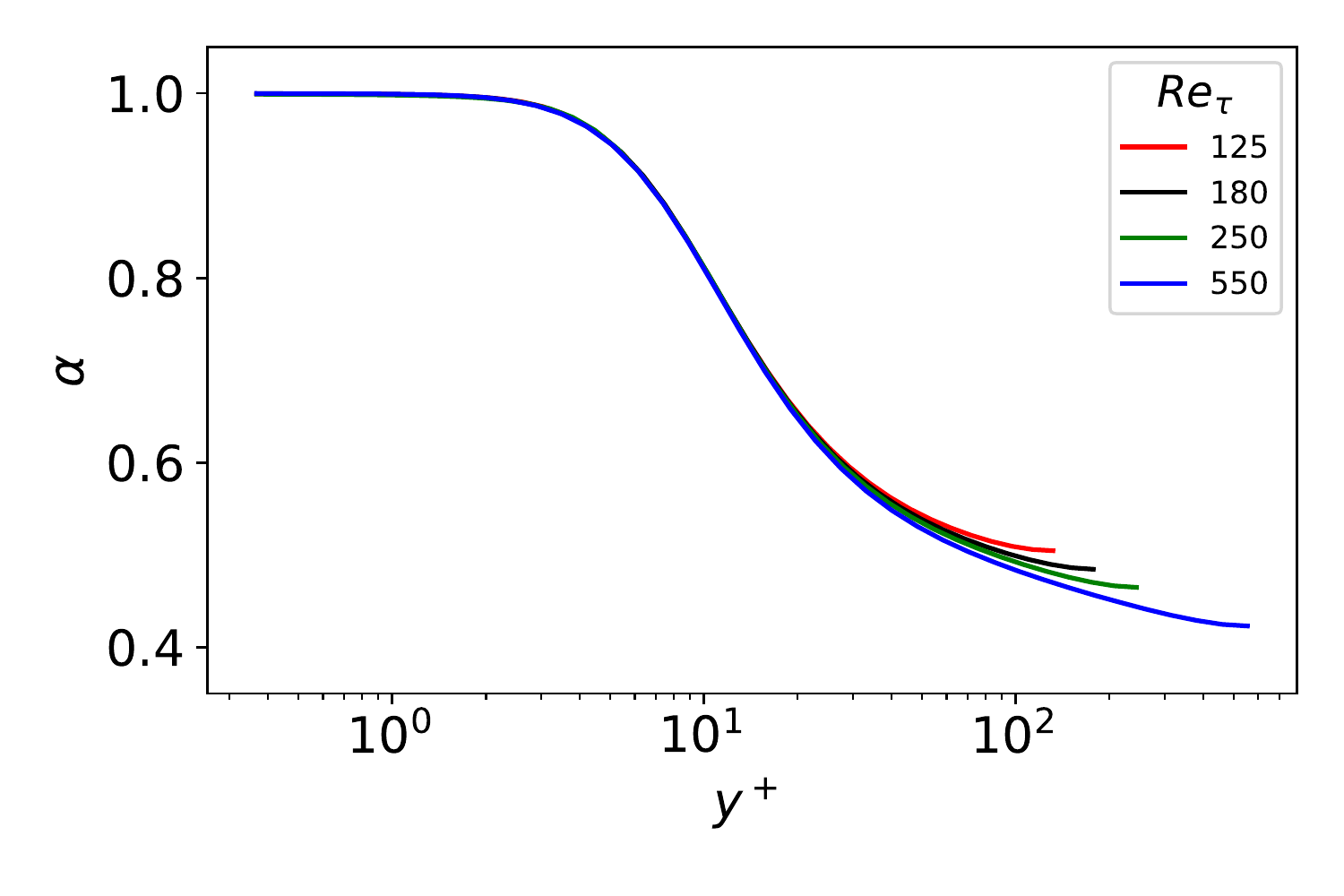}}}%
    \caption{Fractional order of Couette flow for $Re_\tau = 125, 180, 250, 550$ shown for both  one-sided and two-sided model}
    \label{fig:couette_fracorder}
\end{figure}

\begin{figure}[H]
\centering
    \subfloat[one-sided]{{\includegraphics[trim=0.3cm 0.3cm 0.3cm 0.3cm,clip, width=0.5\textwidth]{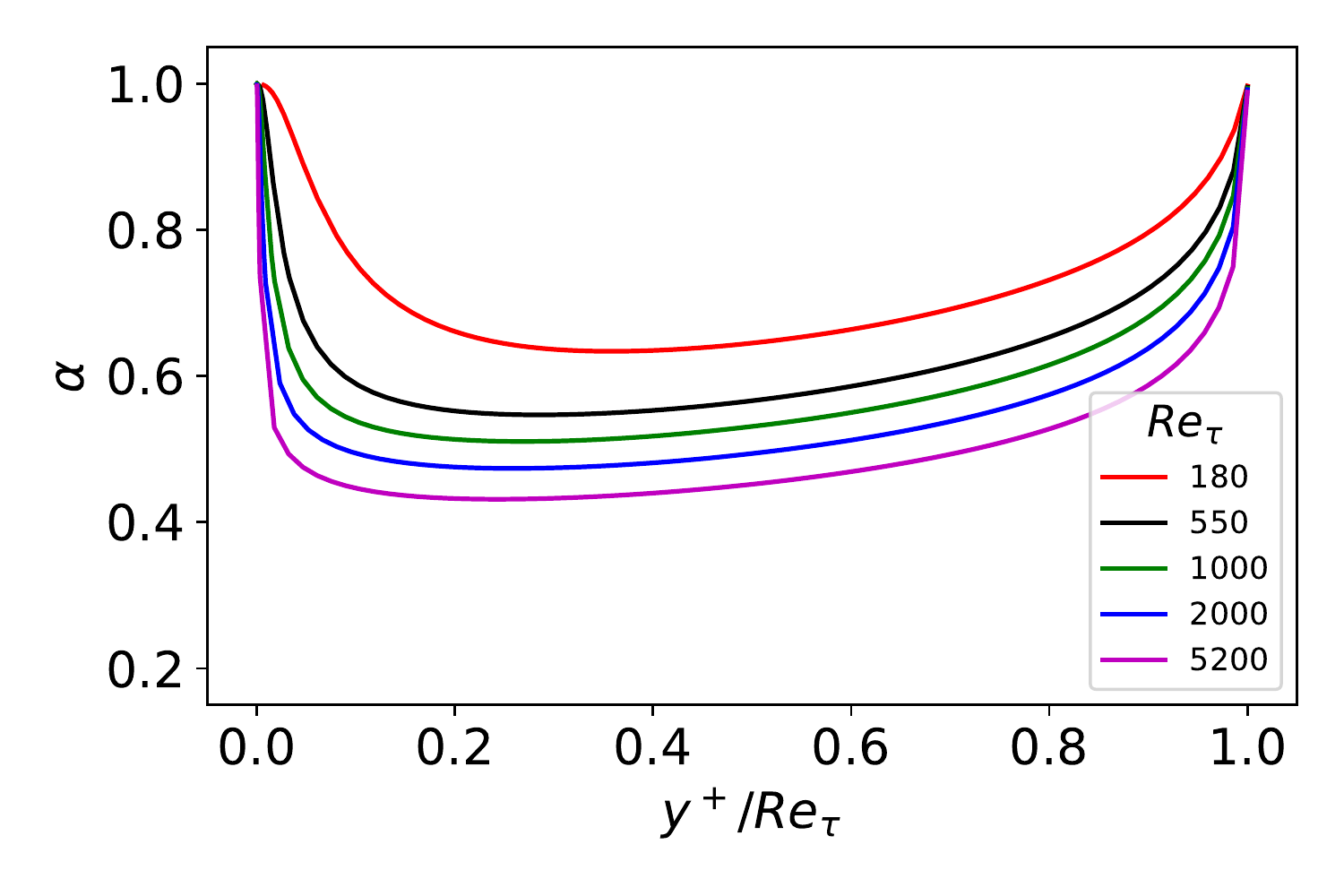}}}%
   \subfloat[two-sided]{{\includegraphics[trim=0.3cm 0.3cm 0.3cm 0.3cm,clip,    width=0.5\textwidth]{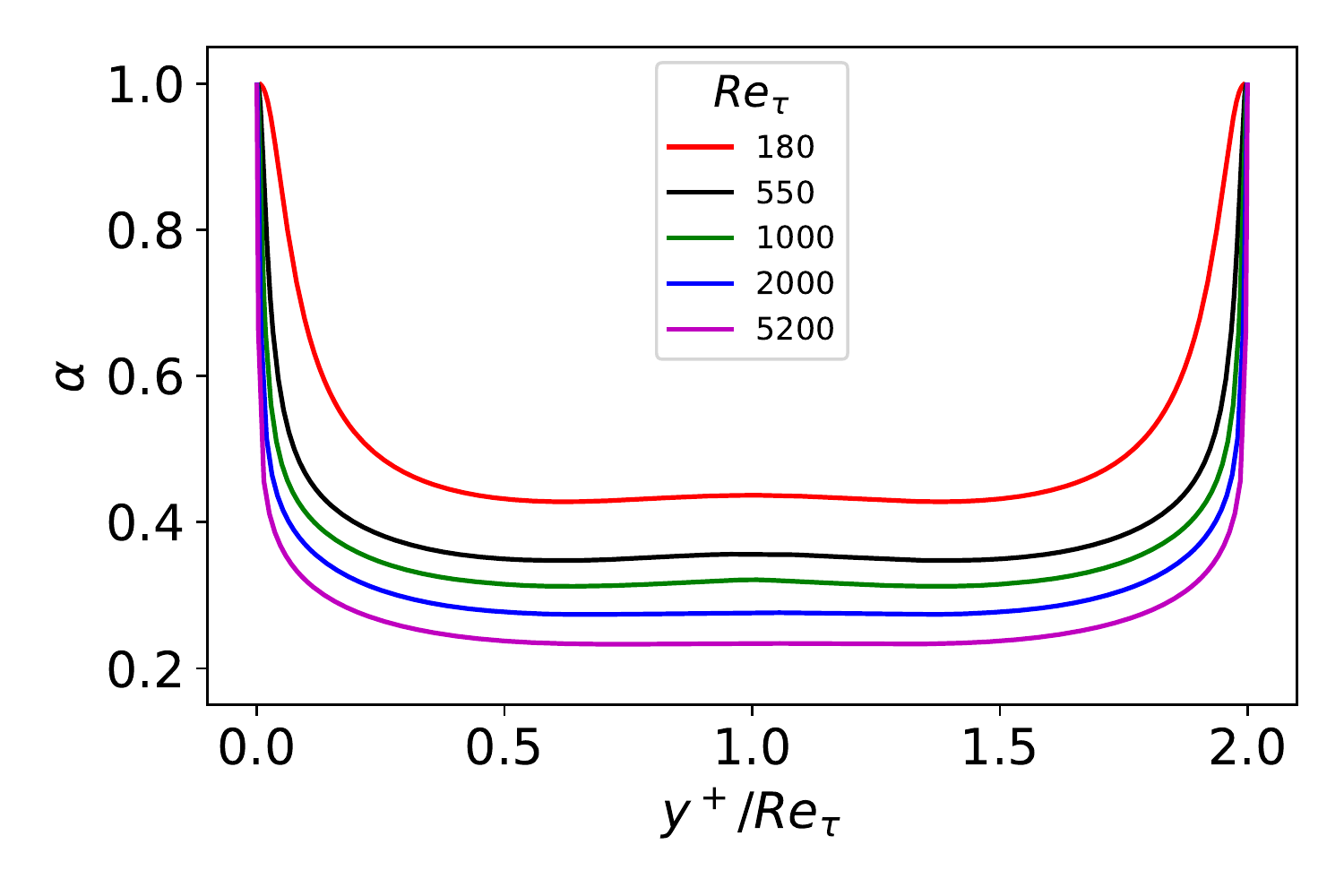}}} \\
   \subfloat[one-sided]{{\includegraphics[trim=0.3cm 0.3cm 0.3cm 0.3cm,clip, width=0.5\textwidth]{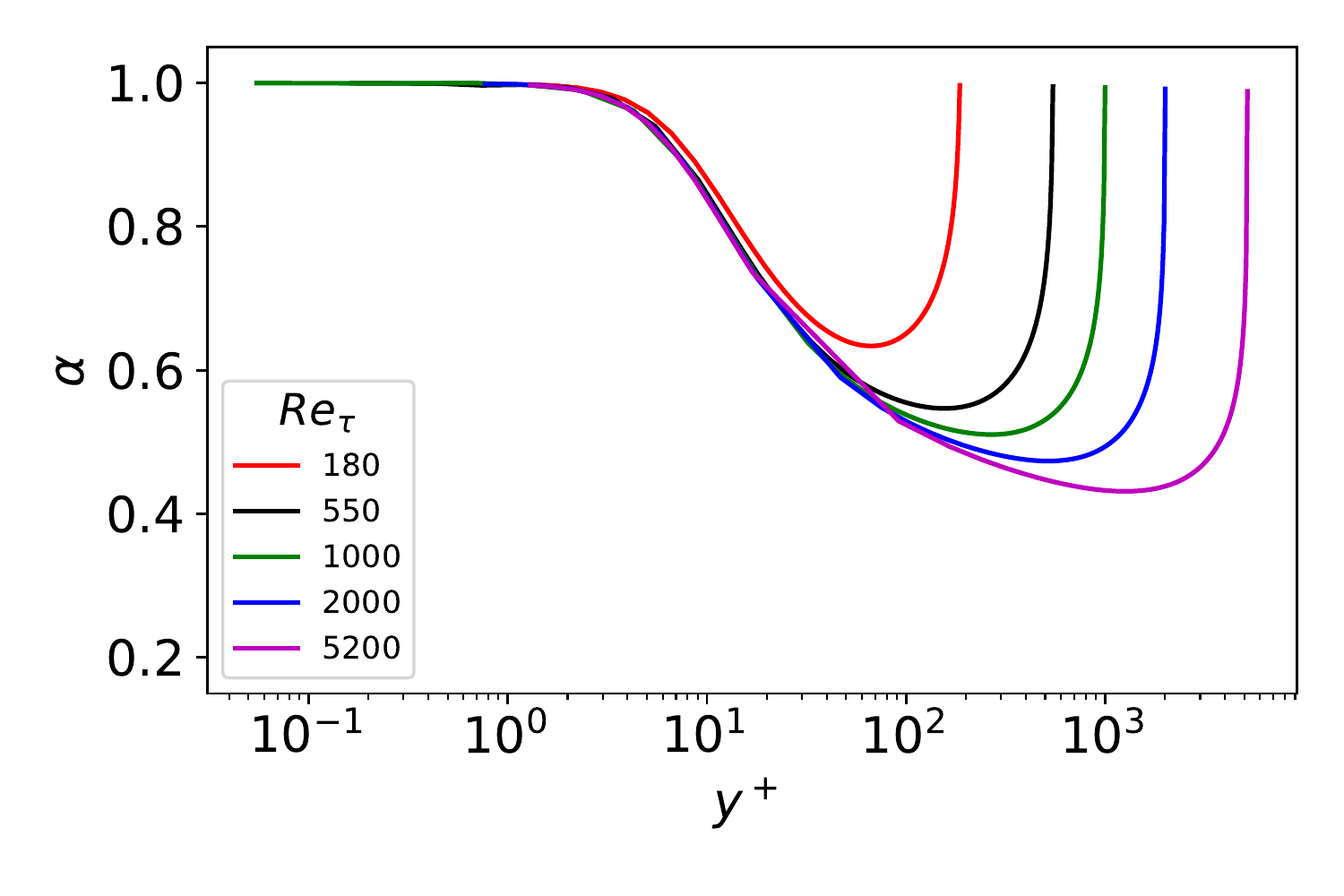}}}%
   \subfloat[two-sided]{{\includegraphics[trim=0.3cm 0.3cm 0.3cm 0.3cm,clip,    width=0.5\textwidth]{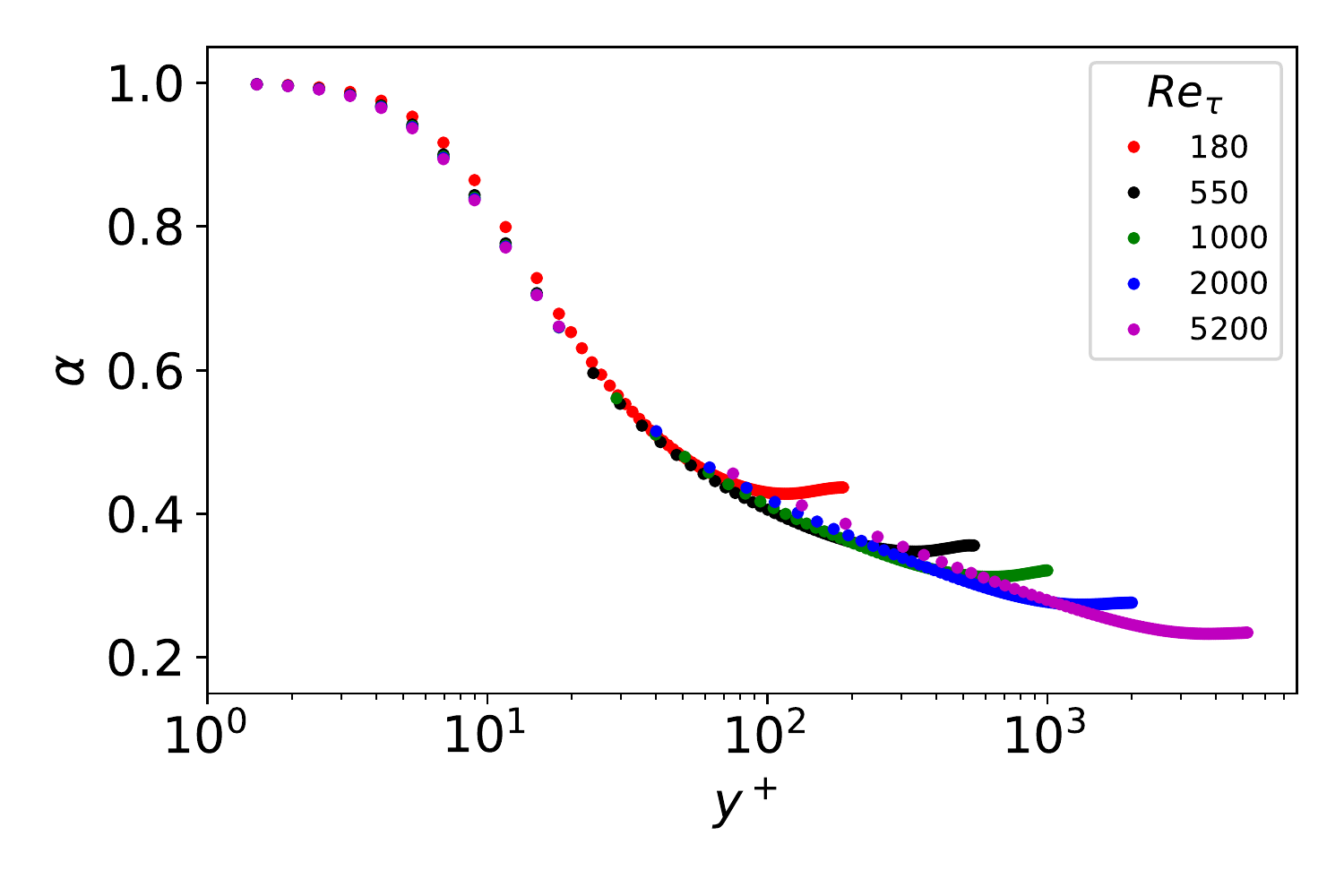}}}%
    \caption{Fractional order of Channel flow for $Re_\tau = 180, 550, 1000, 2000, 5200$ shown for both  one-sided and two-sided model }
    \label{fig:channel_fracorder}
\end{figure}

\begin{figure}[H]
\centering
    \subfloat[one-sided]{{\includegraphics[trim=0.3cm 0.3cm 0.3cm 0.3cm,clip, width=0.5\textwidth]{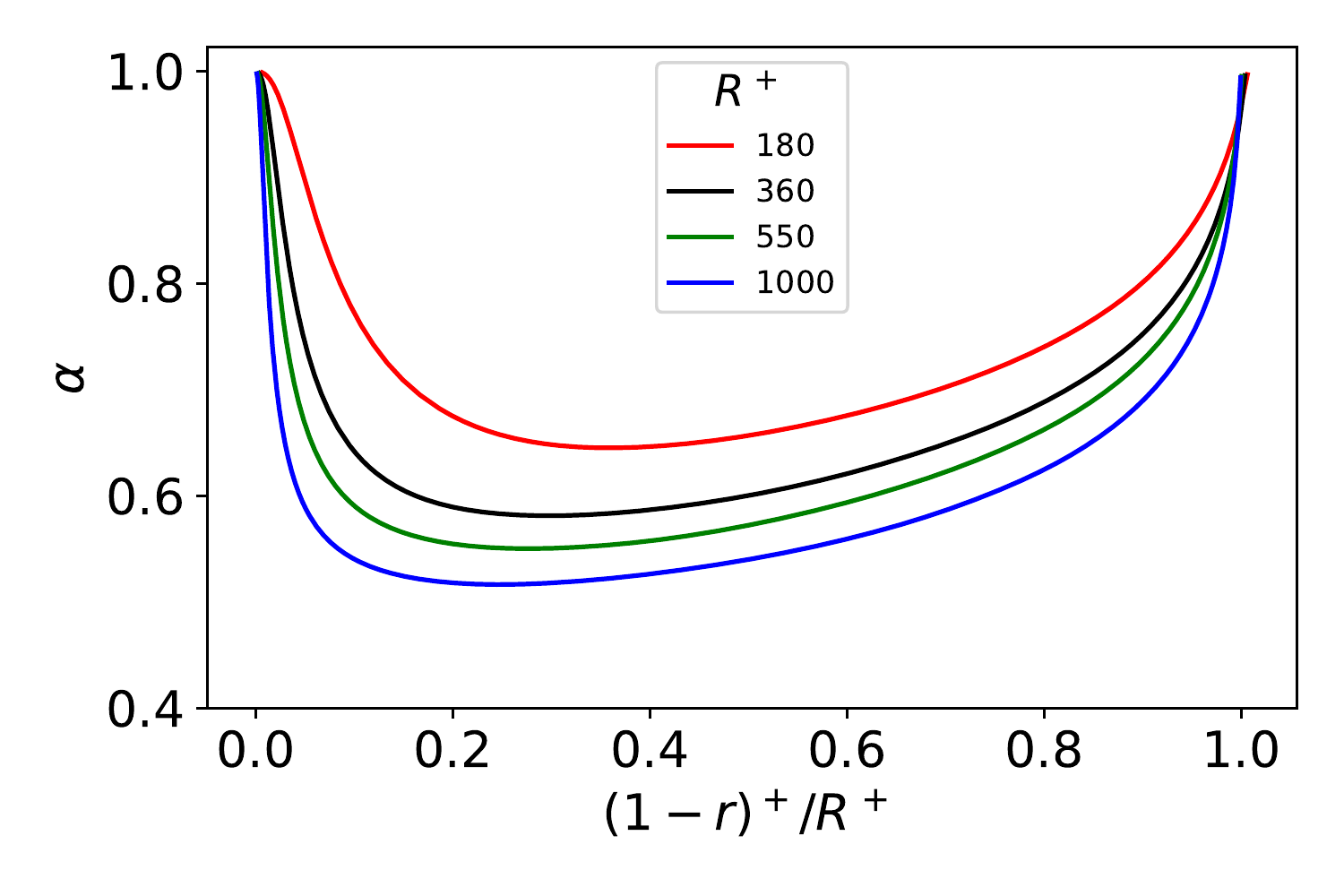}}}%
   \subfloat[two-sided]{{\includegraphics[trim=0.3cm 0.3cm 0.3cm 0.3cm,clip,    width=0.5\textwidth]{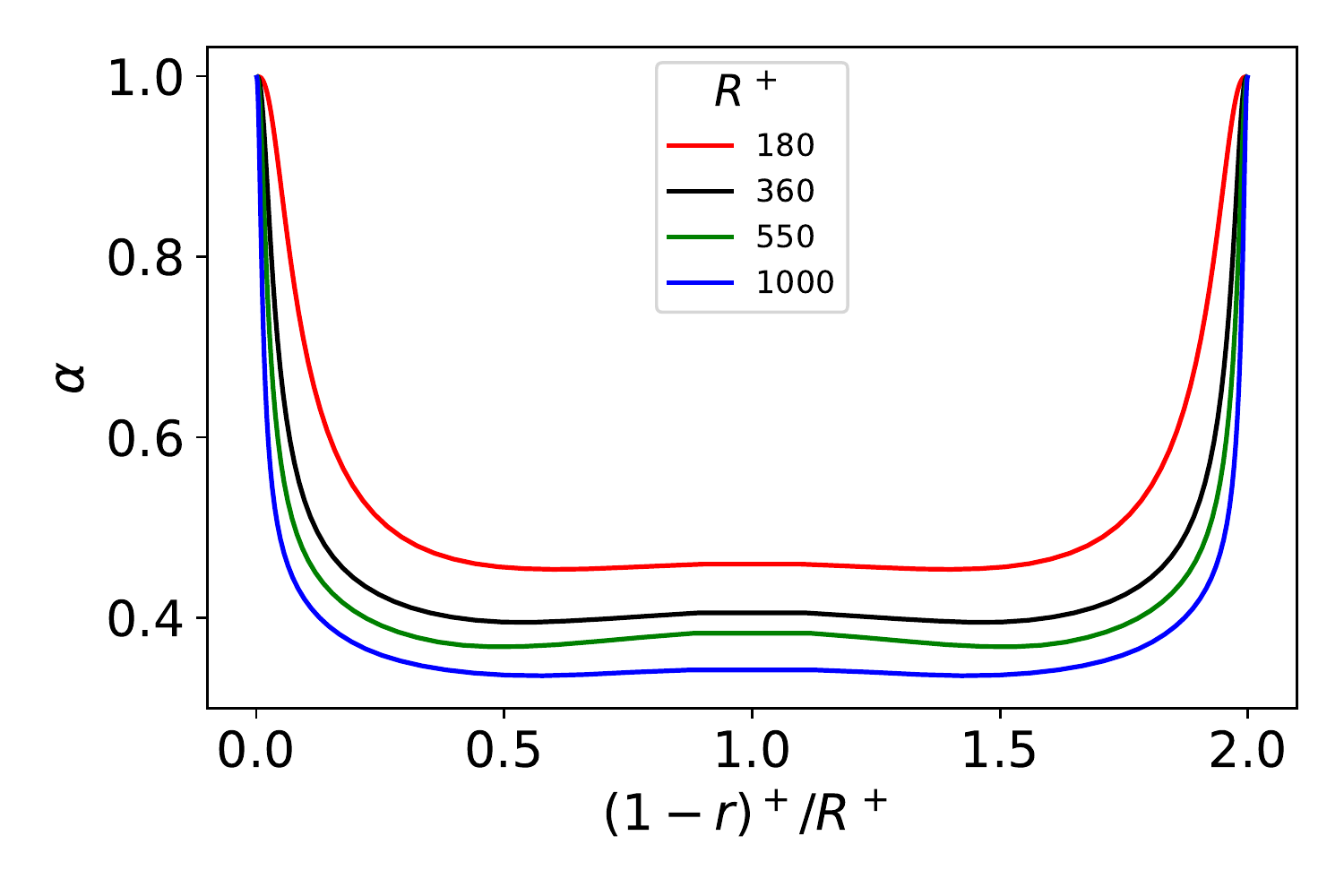}}} \\
   \subfloat[one-sided]{{\includegraphics[trim=0.3cm 0.3cm 0.3cm 0.3cm,clip, width=0.5\textwidth]{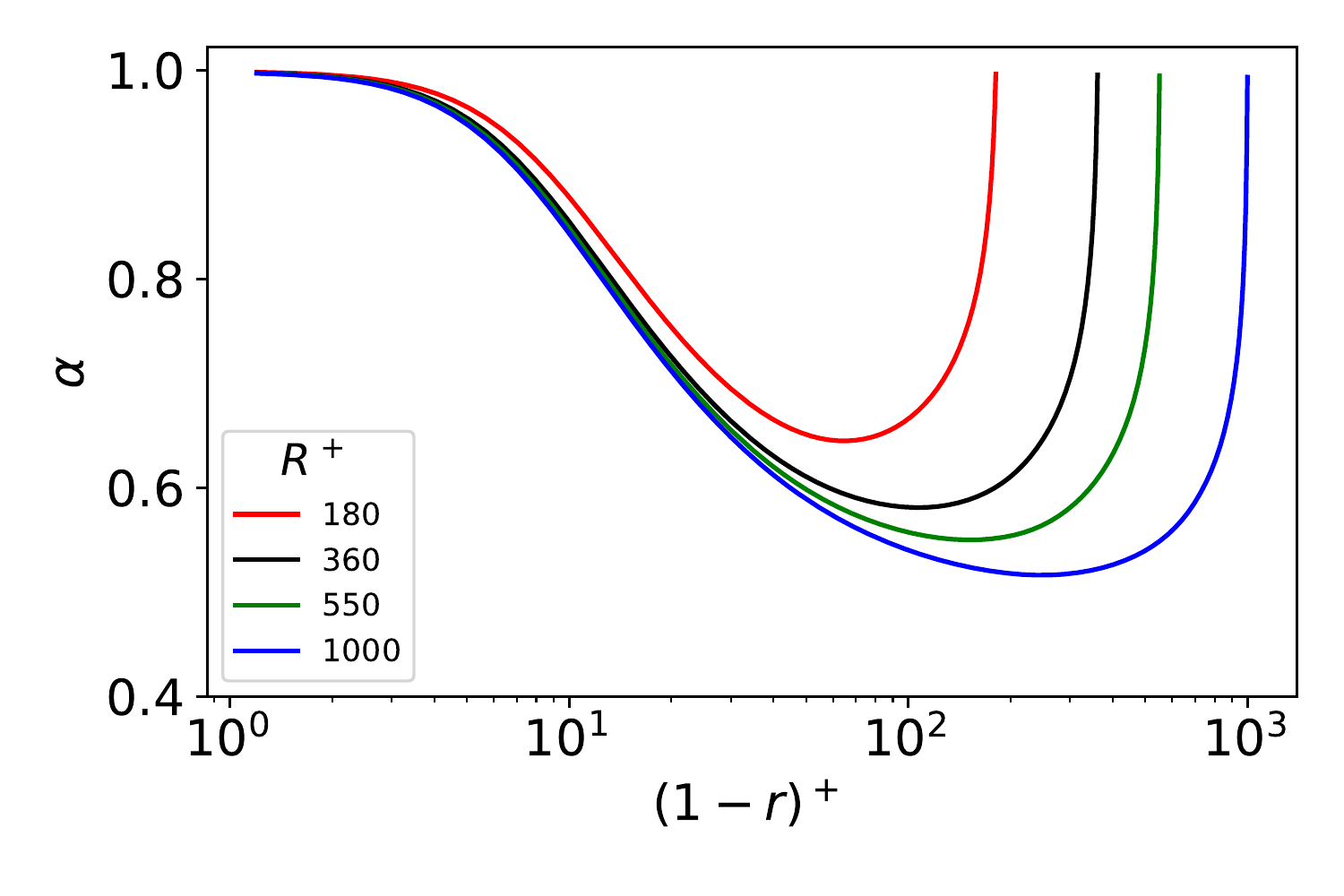}}}%
   \subfloat[two-sided]{{\includegraphics[trim=0.3cm 0.3cm 0.3cm 0.3cm,clip,    width=0.5\textwidth]{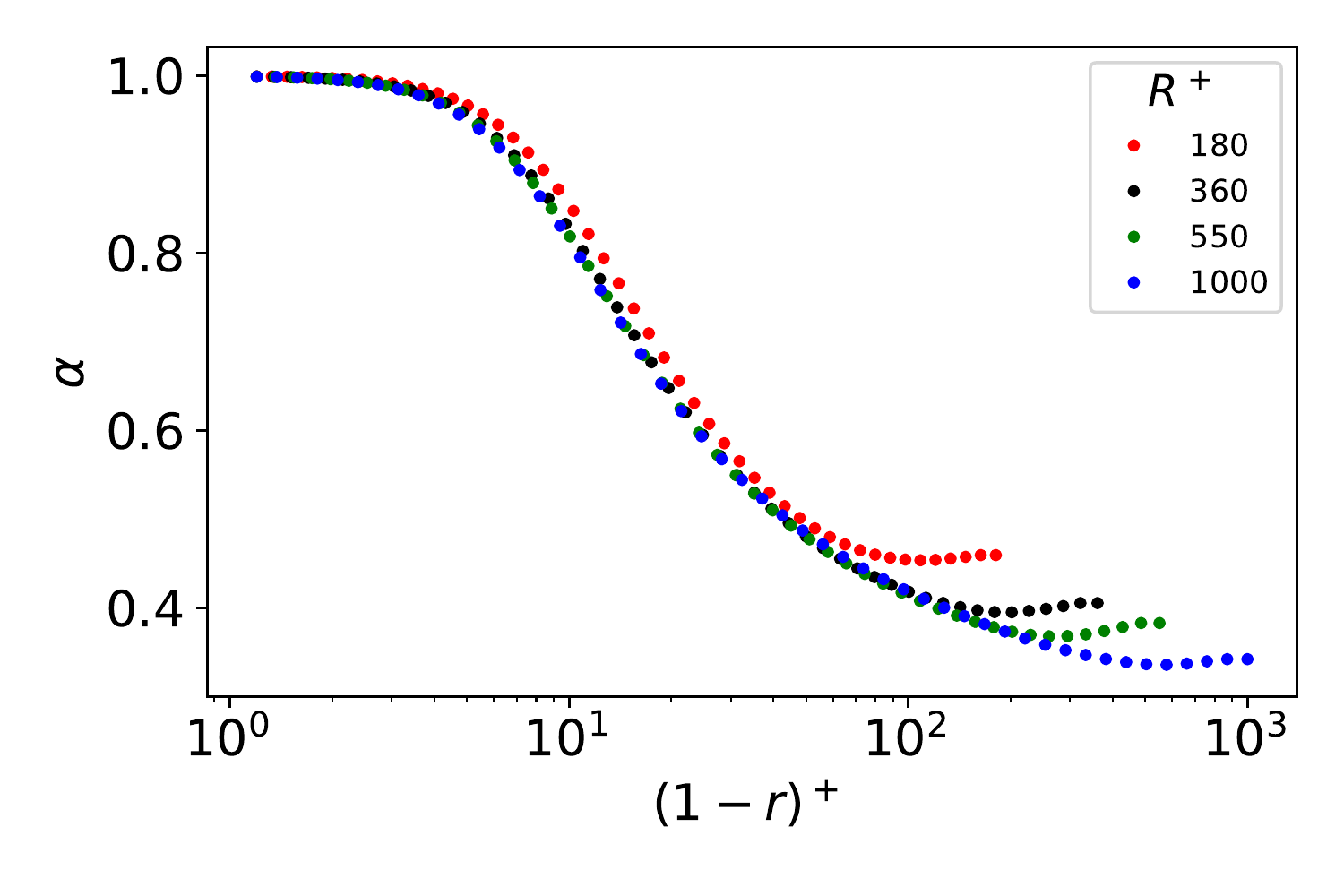}}}%
    \caption{Fractional order of Pipe flow for $Re_\tau = 180, 360, 550, 1000$ shown for both  one-sided and two-sided model }
    \label{fig:pipe_fracorder}
\end{figure}

Inferring from fig.~\ref{fig:couette_fracorder}, \ref{fig:channel_fracorder} and \ref{fig:pipe_fracorder}, in the viscous dominated region, the fractional order is unity implying that the flow is entirely governed by local operators, such as the near wall region of these three flows for both one- and two-sided model. Furthermore, they exhibit universality as observed in both one- and two- sided formulation in the viscous dominated regions. As we enter the buffer region and eventually to the outer-turbulent regions for these wall bounded flows, the non-locality increases gradually rather than abruptly as indicated by lowering of the fractional order. This is consistent with the physical observations where there is no discontinuity but rather a smooth transition. Thus, it is evident that fluid flow exhibits duality of local and non-local characteristics.

Inferring from fig.~\ref{fig:couette_fracorder}(c), \ref{fig:channel_fracorder}(c) and \ref{fig:pipe_fracorder}(c), only the one-sided formulation for Couette flow exhibits universality, while the one-sided model for Channel and Pipe shows an anomaly at the center-line, where the fractional order is unity. This anomaly is an artifact of the symmetry. The inertial forces are dominant at the center-line with the viscous contribution in close proximity to zero and the total shear stress which includes only the Reynolds stress, which is zero at the center-line due to symmetry, therefore, unity is a numerical solution for the fractional order, however, physically the non-locality in this region is stronger than any other region 

\begin{figure}[!htb]
\centering
    \subfloat{{\includegraphics[trim=0.3cm 0.3cm 0.3cm 0.3cm,clip, width=0.5\textwidth]{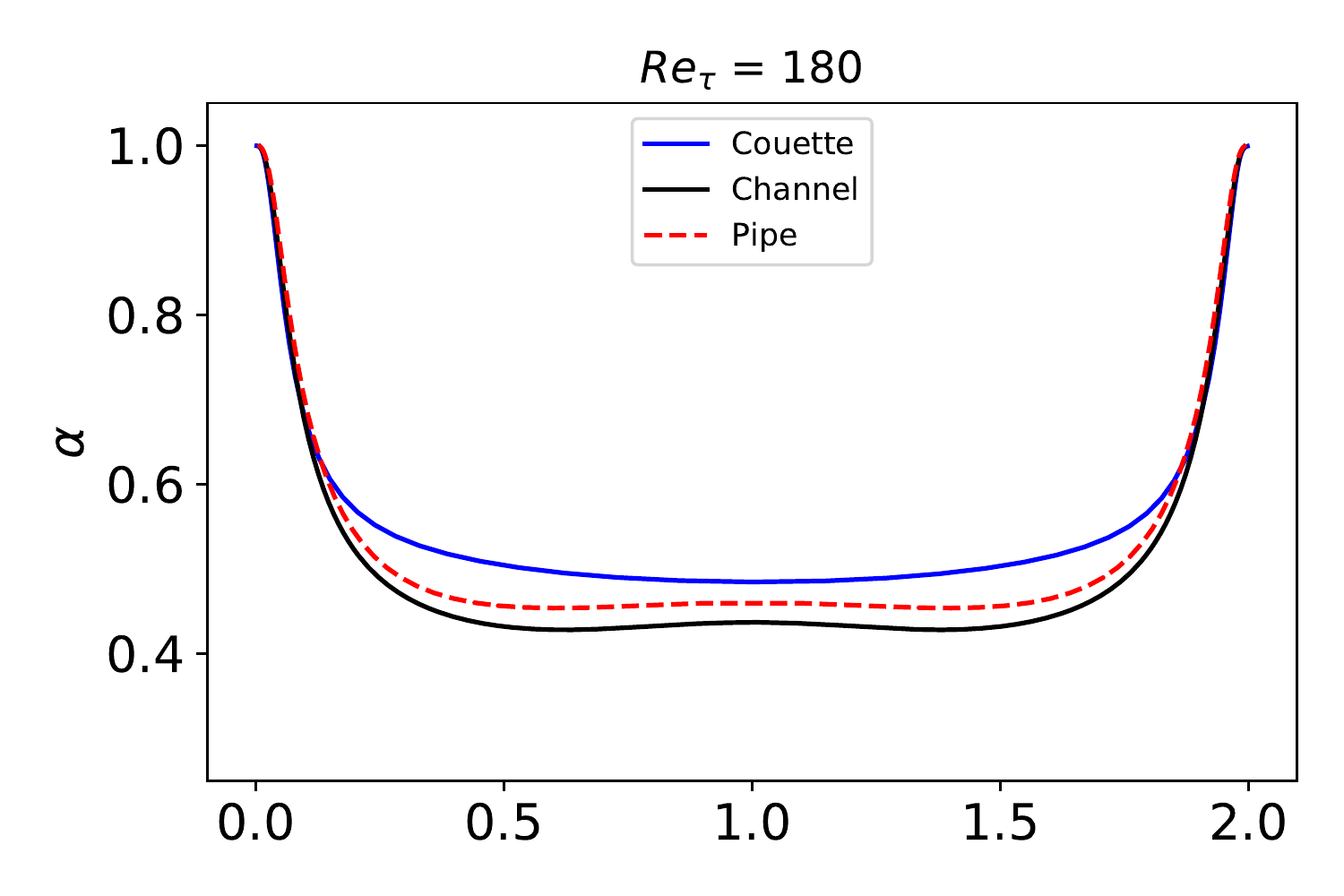}}}%
   \subfloat{{\includegraphics[trim=0.3cm 0.3cm 0.3cm 0.3cm,clip,    width=0.5\textwidth]{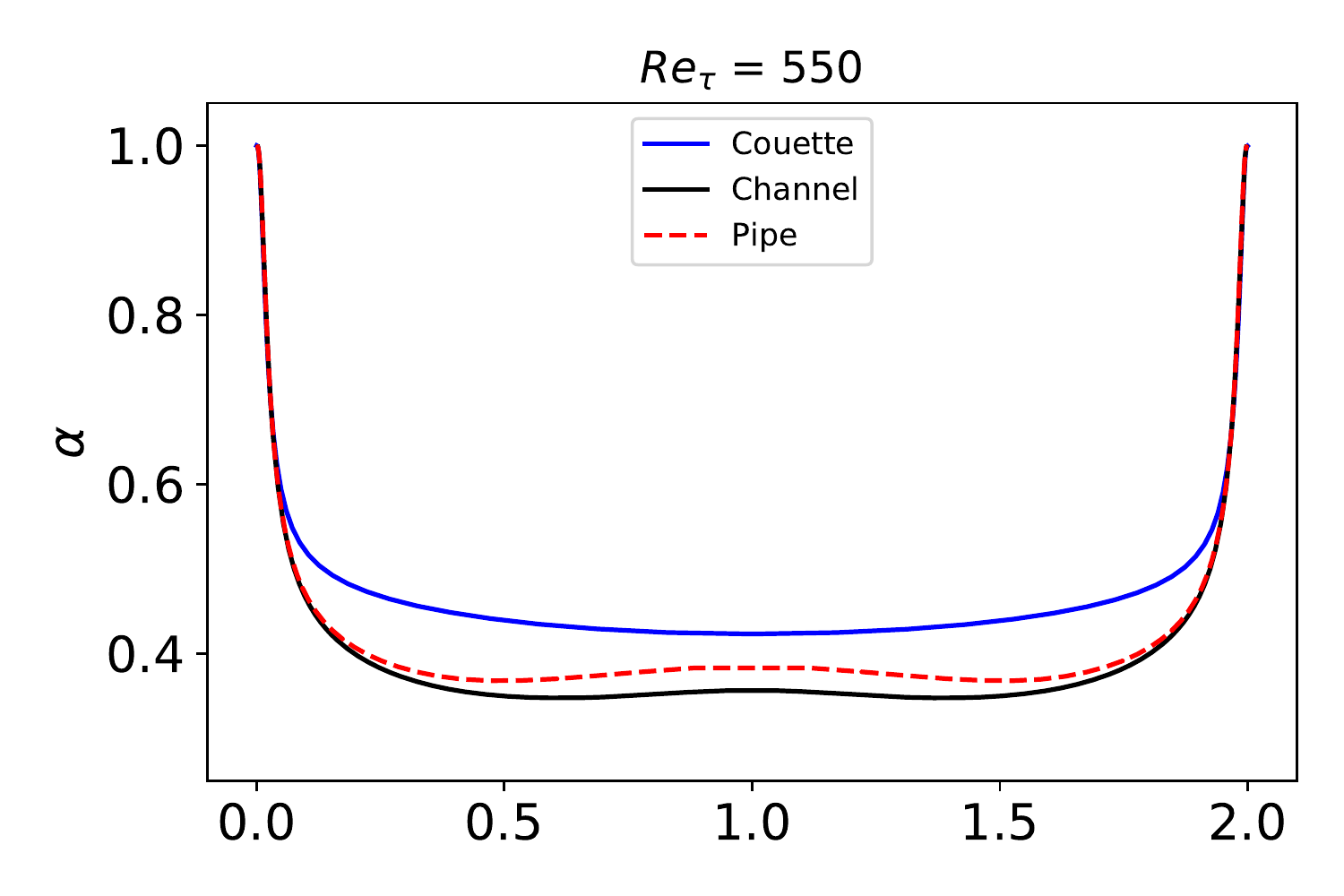}}}%
    \caption{Comparison of fractional order of two-sided f-RANS model for $Re_\tau = 180, 550$ for Channel, Couette and Pipe flow. Here the x-axis of the plot: Channel and Couette: $y^+/Re_\tau$, Pipe: $(1-r)^+/R^+$}
    \label{fig:camp_two_ydel}
\end{figure}

Non-locality at a given point is a result of net disturbances from all the sources, which in these cases are the wall, since they are on both sides, it natural to formulate a two-sided model. The anomaly of the one-sided model (fig.~\ref{fig:couette_fracorder}(c), \ref{fig:channel_fracorder}(c) and \ref{fig:pipe_fracorder}(c)) is no longer seen in the two sided model (fig.~\ref{fig:couette_fracorder}(d), \ref{fig:channel_fracorder}(d) and \ref{fig:pipe_fracorder}(d)), where the result are physically consistent. However, as a result of the symmtery, the two-sided model has a discontinuity at the centerline for the case of Channel and Pipe, where all values of fractional order, $\alpha \in (0, 1]$ is a solution numerically. Thus in the plots (fig.~\ref{fig:channel_fracorder}(b), \ref{fig:pipe_fracorder}(b) and \ref{fig:camp_two_ydel}), the average value of the two neighbours is used to maintain the smoothness of the curve. Furthermore, as the Reynolds number increases the flow is characterised by higher turbulence intensity, which corresponds to a lower fractional order in fig.~\ref{fig:couette_fracorder}(b,d), \ref{fig:channel_fracorder}(b,d) and \ref{fig:pipe_fracorder}(b,d). 

We further compare the fractional order for these three cases for approximately the same friction Reynolds number (fig.~\ref{fig:camp_two_ydel}). The Channel and Pipe shows a "bump" near the center line, which is not observed in the case of Couette. It is to be noted as the friction Reynolds number increases $\approx 1000$ this "bump" flattens. The reason of this "bump" is unknown and its is to be believed as an effect of symmetry.

\begin{figure}[H]
\centering
    \subfloat[Couette]{{\includegraphics[trim=0.3cm 0.3cm 0.3cm 0.3cm,clip, width=0.33\textwidth]{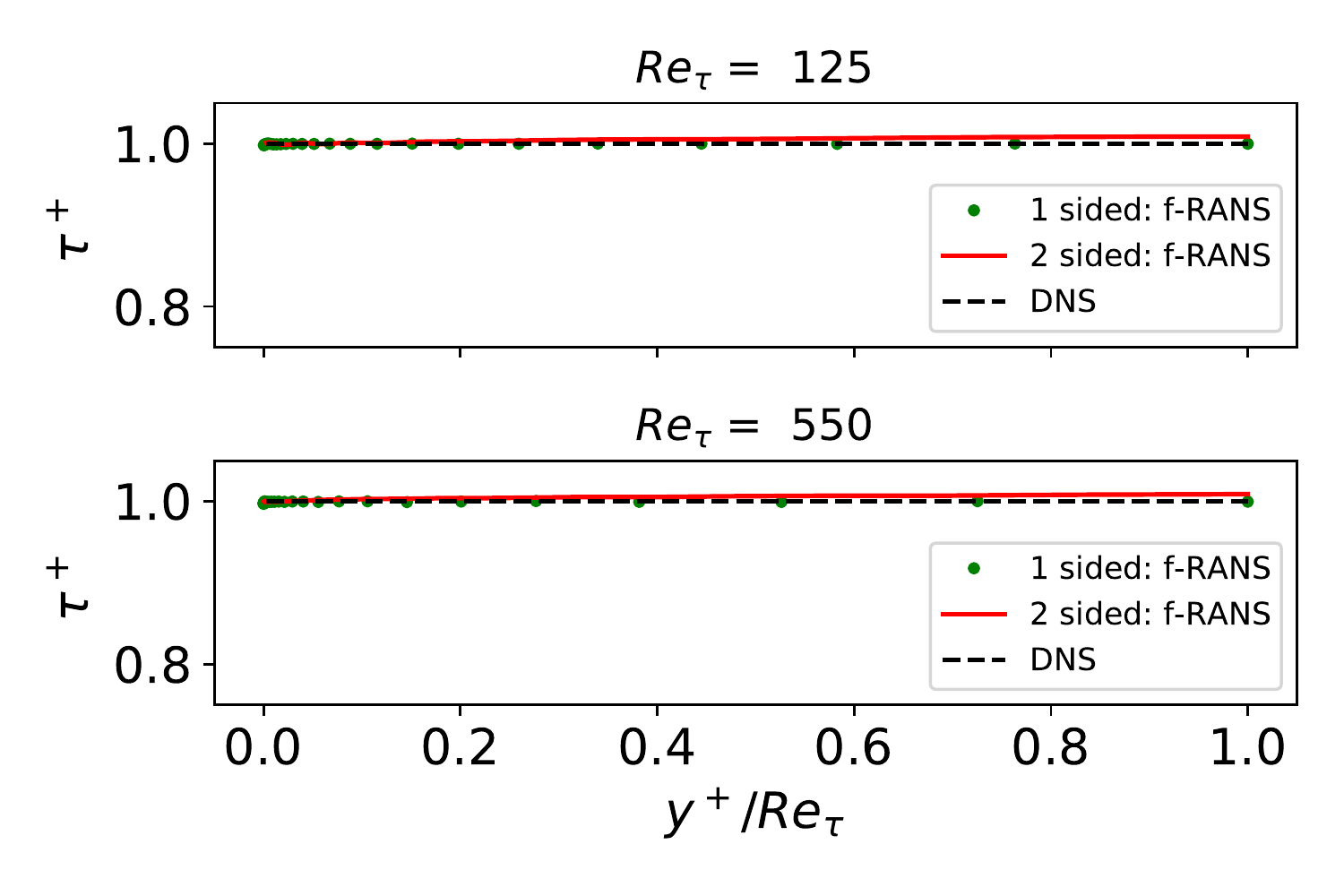}}}%
   \subfloat[Channel]{{\includegraphics[trim=0.3cm 0.3cm 0.3cm 0.3cm,clip,    width=0.33\textwidth]{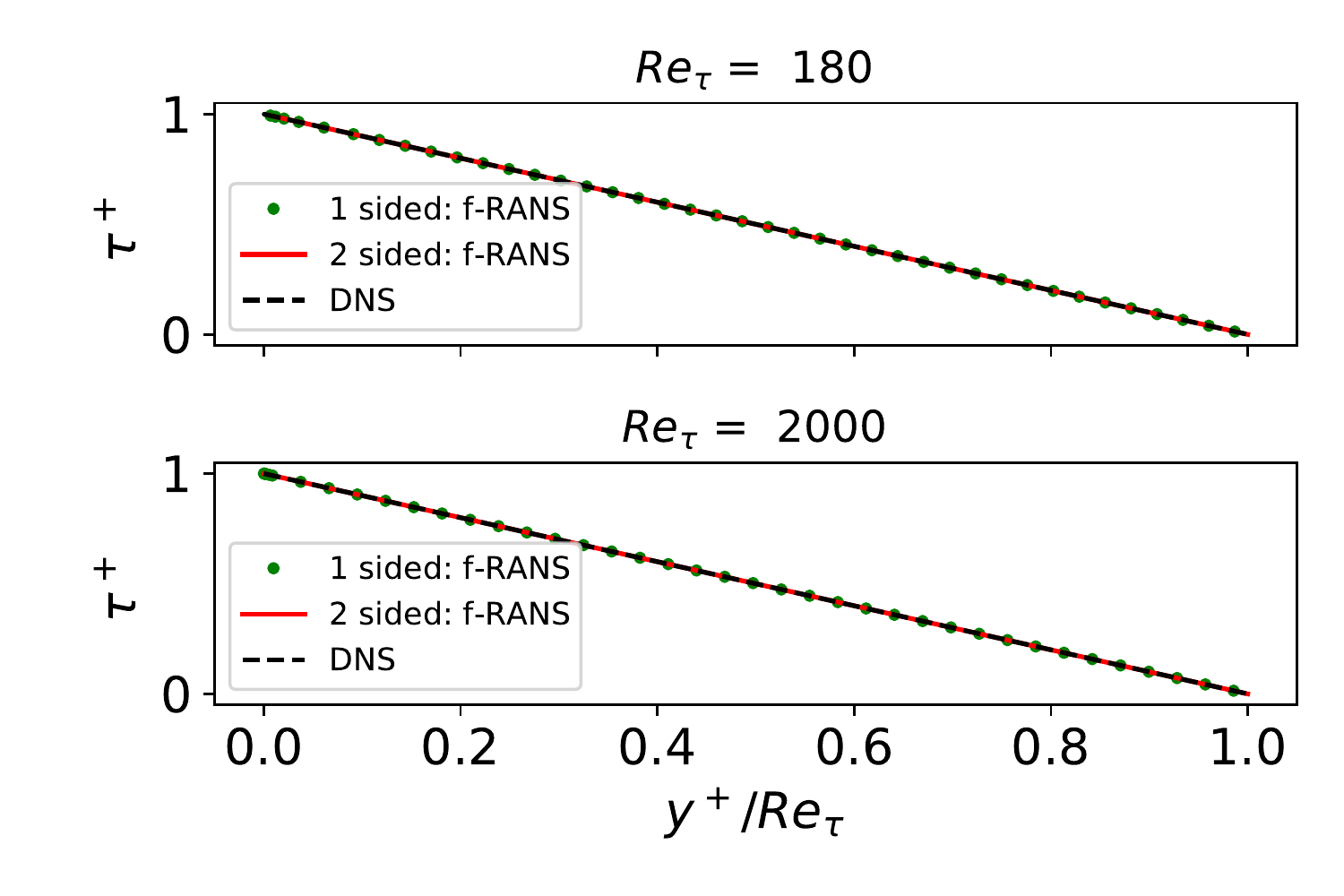}}}%
    \subfloat[Pipe]{{\includegraphics[trim=0.3cm 0.3cm 0.3cm 0.3cm,clip,    width=0.33\textwidth]{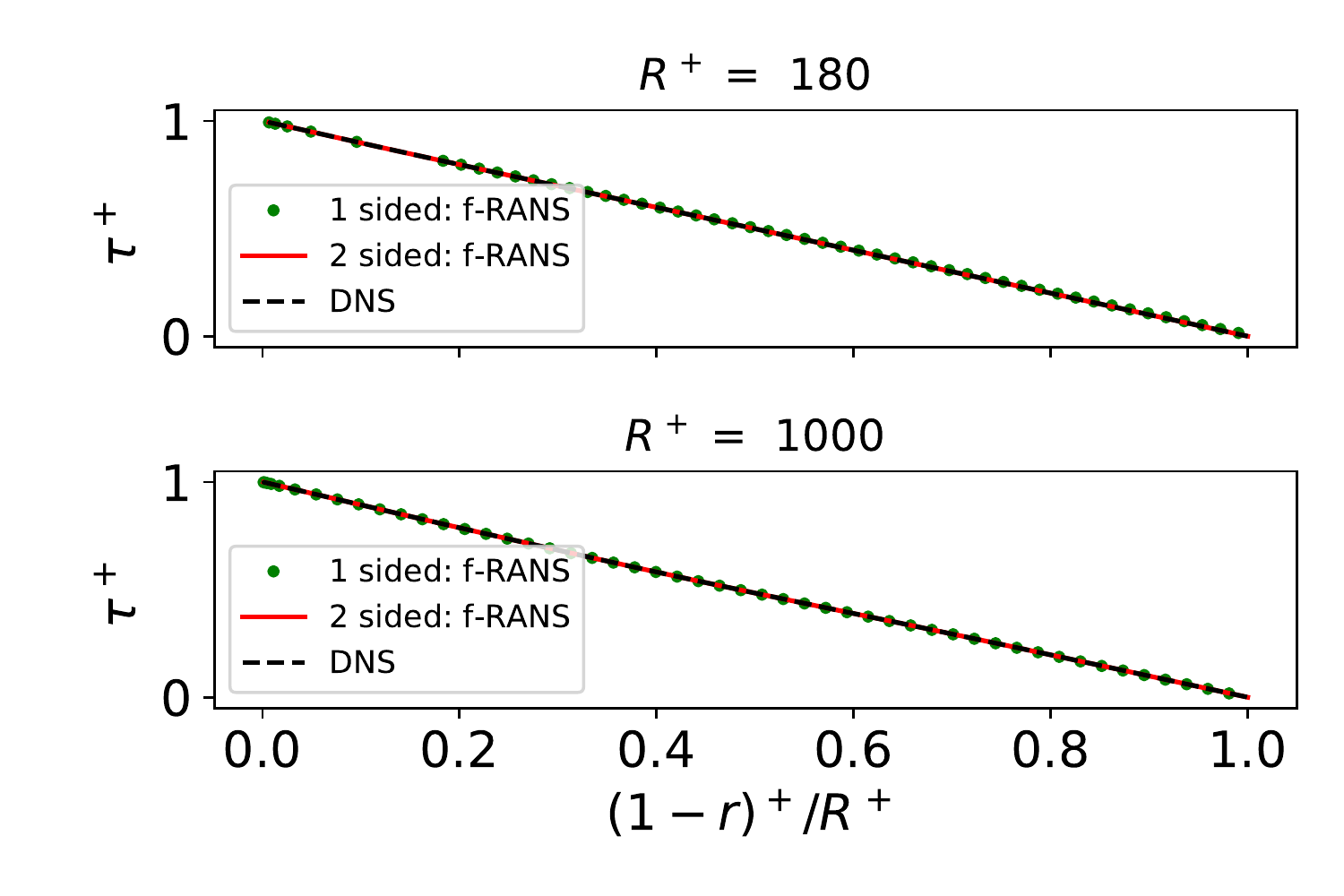}}}%
    \caption{Comparison of total Shear Stress obtained using one- and two-sided f-RANS model with DNS databases for Couette, Channel and Pipe flow}
    \label{fig:tau_cc}
\end{figure}

\begin{figure}[H]
\centering
    \subfloat[Couette]{{\includegraphics[trim=0.3cm 0.3cm 0.3cm 0.3cm,clip, width=0.33\textwidth]{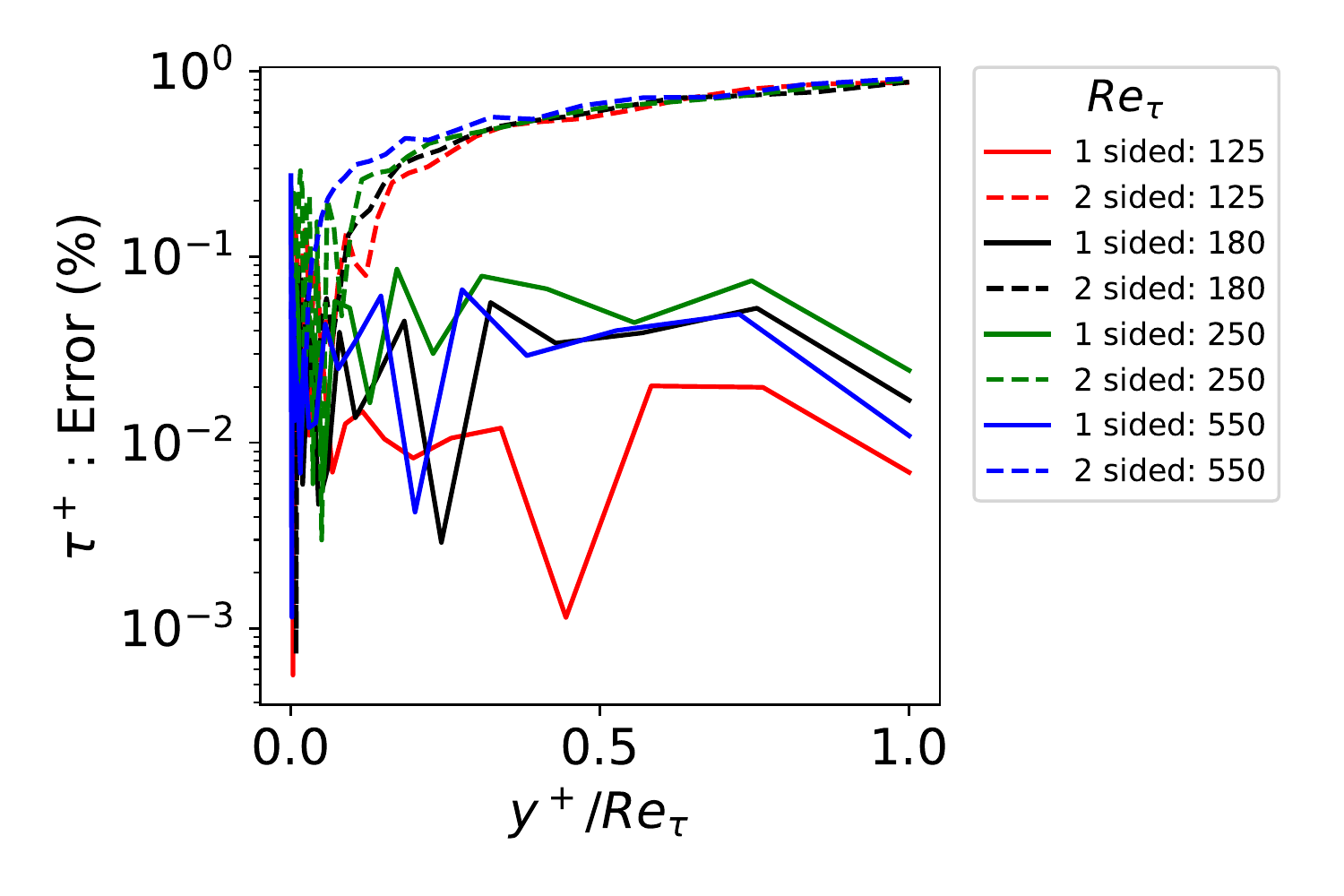}}}%
   \subfloat[Channel]{{\includegraphics[trim=0.3cm 0.3cm 0.3cm 0.3cm,clip,    width=0.33\textwidth]{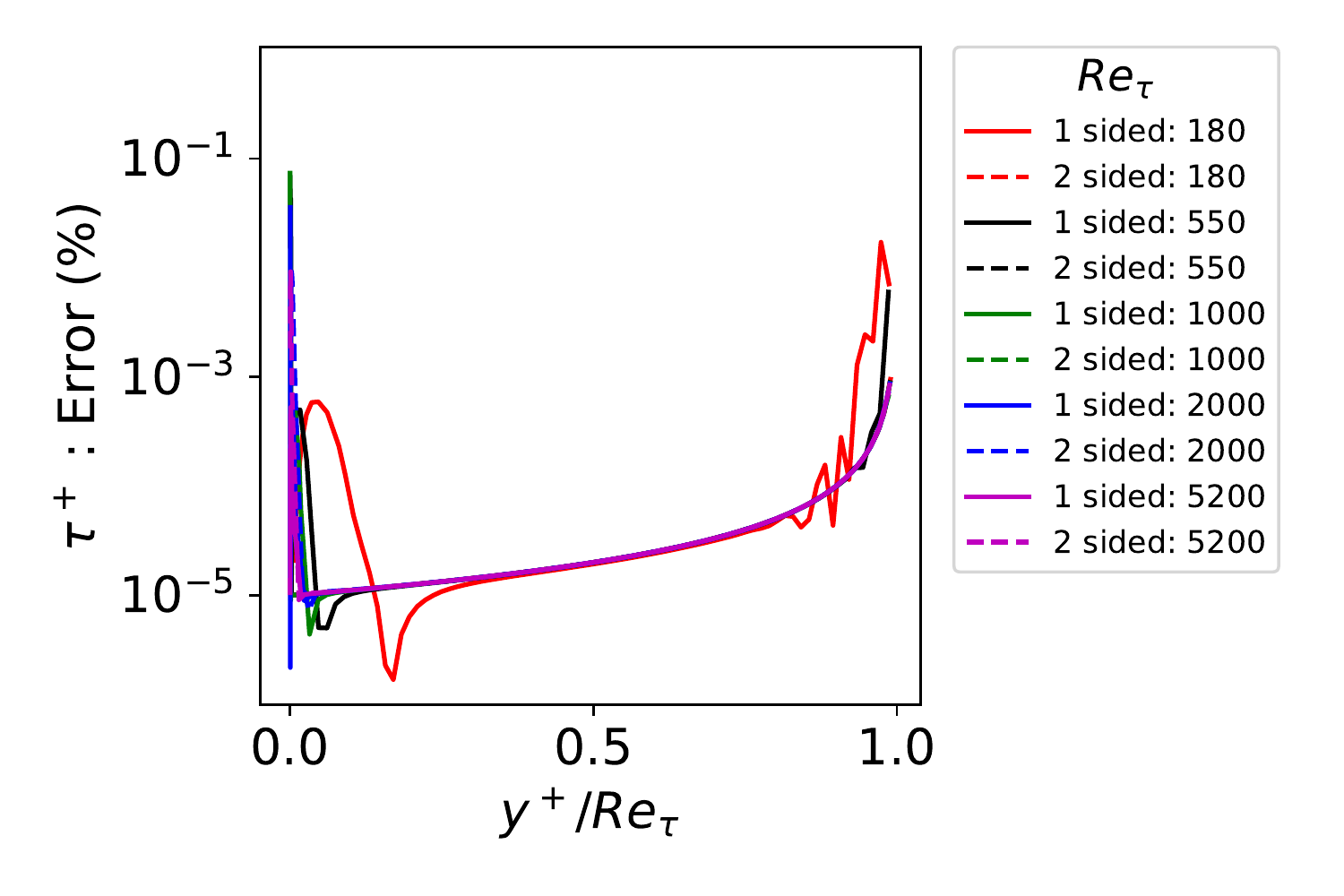}}}%
   \subfloat[Pipe]{{\includegraphics[trim=0.3cm 0.3cm 0.3cm 0.3cm,clip,    width=0.33\textwidth]{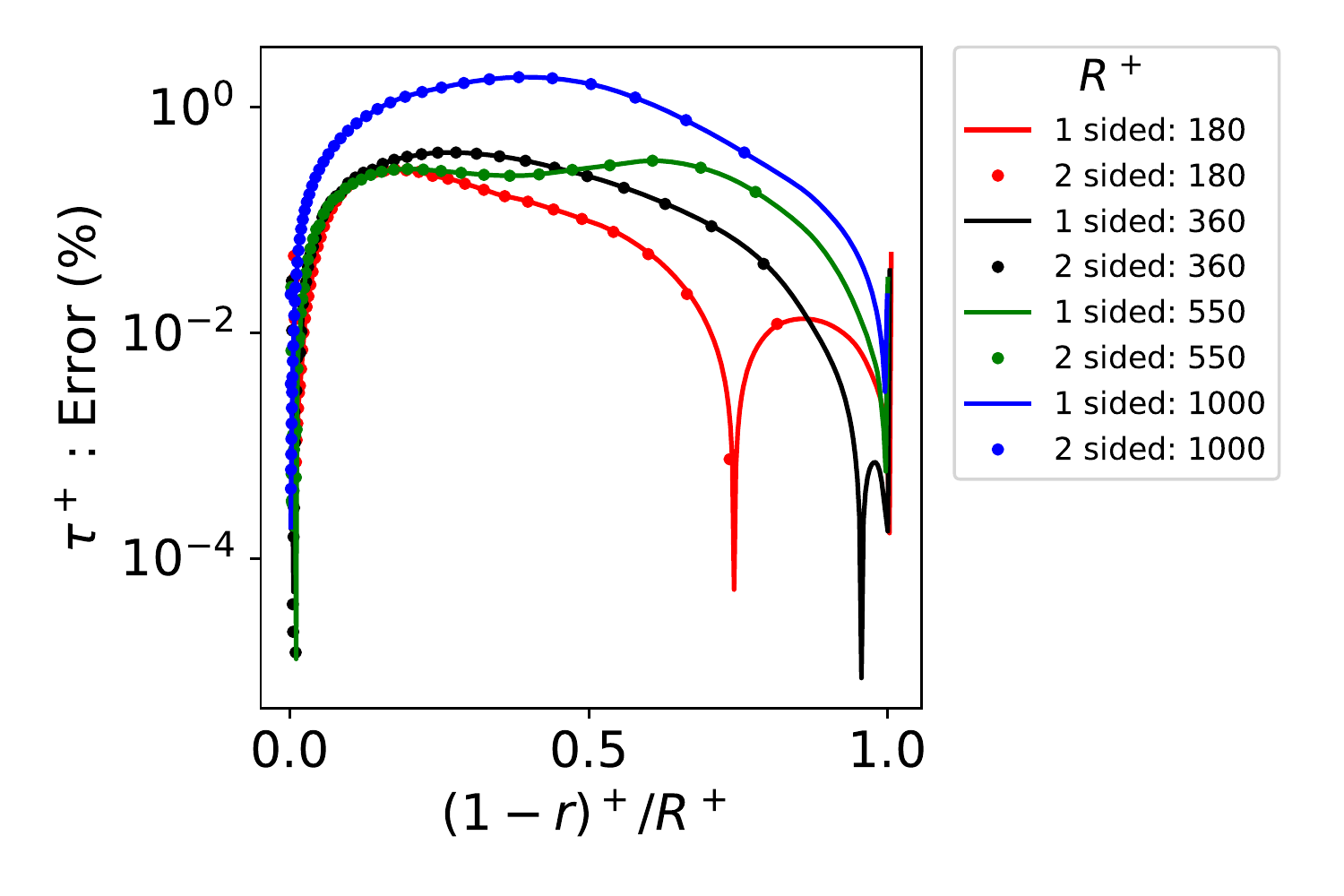}}}%
    \caption{Error of total Shear Stress obtained using one- and two-sided f-RANS model with DNS databases for Couette, Channel and Pipe flow}
    \label{fig:tau_cc_err}
\end{figure}

Finally, we note from fig.~\ref{fig:tau_cc} and \ref{fig:tau_cc_err} that the fractional order computed for either one- or two- sided model gives an error less than $1 \%$ in total shear stress.


\subsection{Asymptotic relationship and scaling laws}

\begin{figure}[!htb]
\centering
    \subfloat[Couette]{{\includegraphics[trim=0.3cm 0.3cm 0.3cm 0.3cm,clip, width=0.33\textwidth]{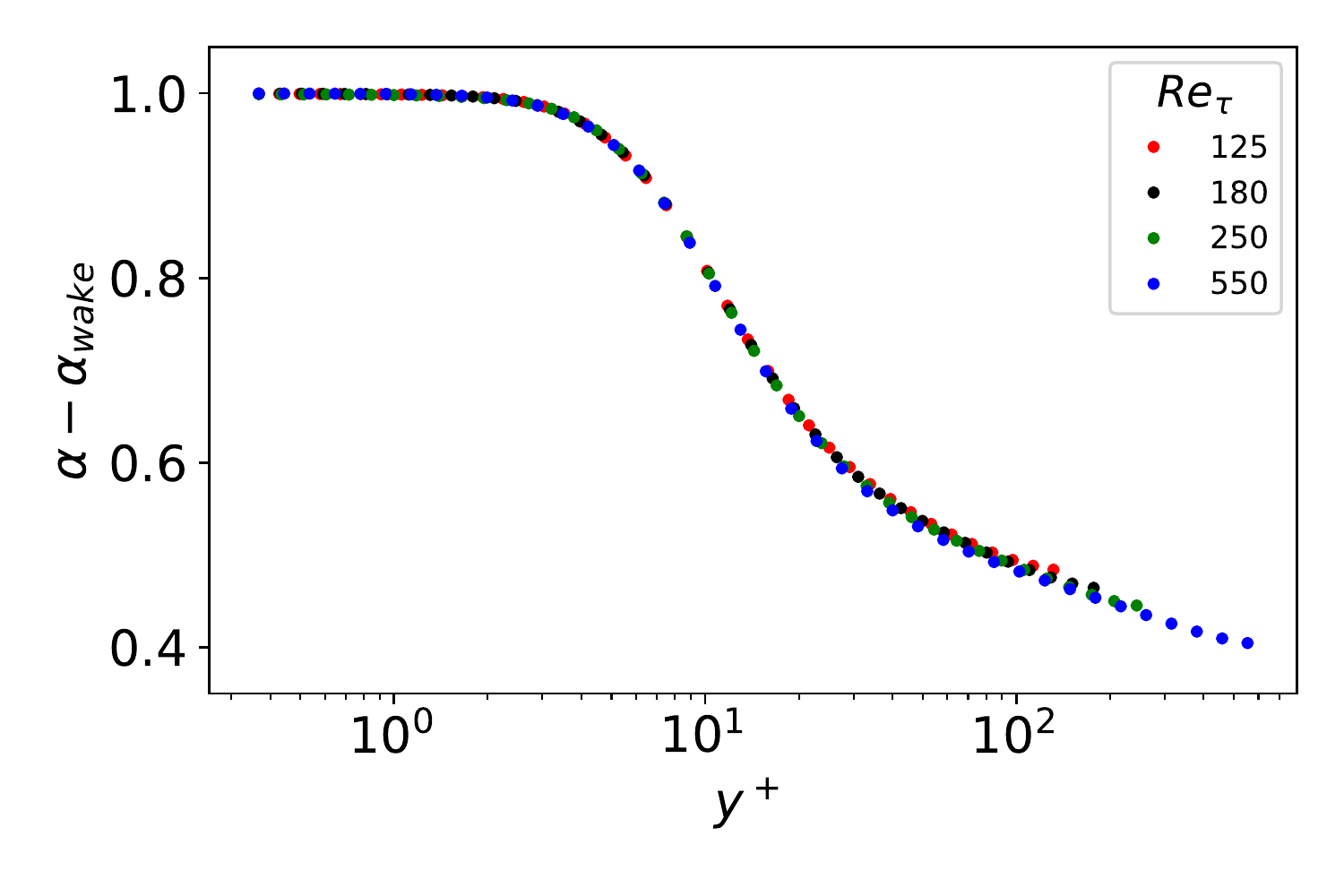}}}%
   \subfloat[Channel]{{\includegraphics[trim=0.3cm 0.3cm 0.3cm 0.3cm,clip,    width=0.33\textwidth]{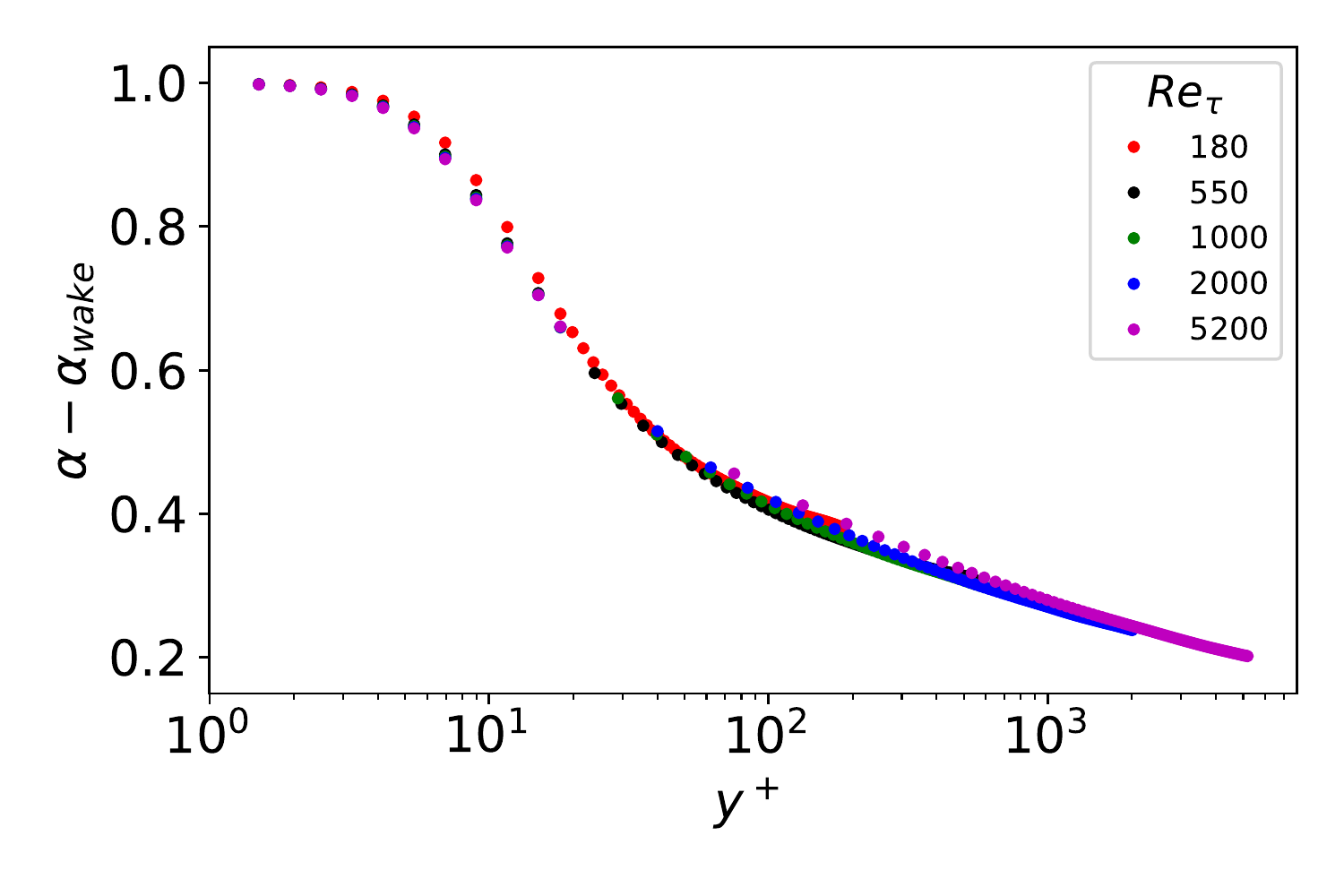}}}%
    \subfloat[Pipe]{{\includegraphics[trim=0.3cm 0.3cm 0.3cm 0.3cm,clip,    width=0.33\textwidth]{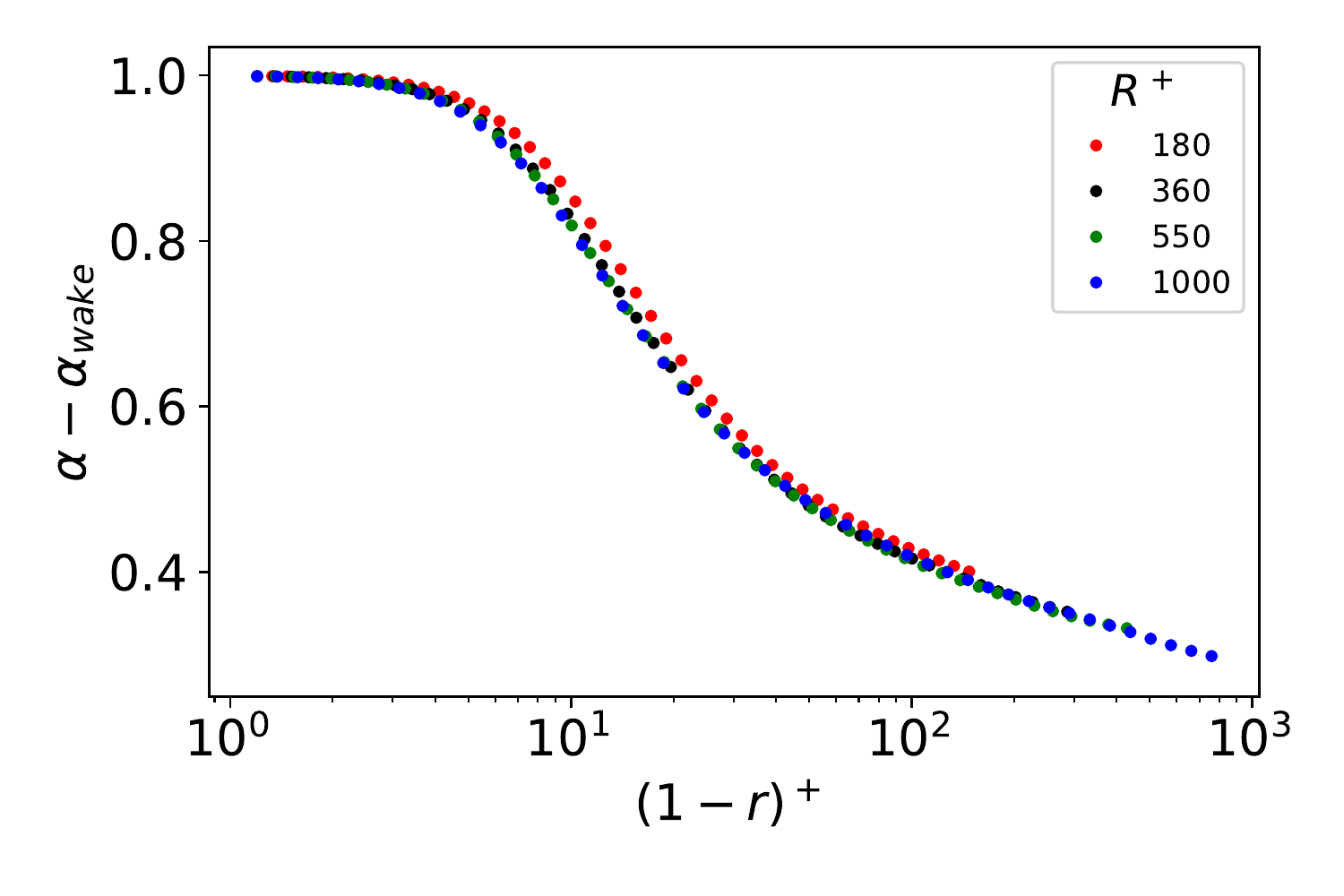}}}%
    \caption{Fraction order for two-sided model shown for the universal part of the curve obtained by subtracting the wake contribution}
    \label{fig:alp_wake}
\end{figure}

Upon close observation in the log-scale of fig.~\ref{fig:couette_fracorder}(d), \ref{fig:channel_fracorder}(d) and \ref{fig:pipe_fracorder}(d) a notion of universality can be seen in the two-sided model, where in the wake region a non-universal scaling is present. We decompose the universal and the wake parts of the curve as (\ref{eq:wake1}). The universal component is a function purely of $y^+$ or $(1-r)^+$, however, the wake region which is a function of $y^+ / Re_\tau$ or $(1-r)^+/R^+$. An analytical expression is deduced by fitting the fractional order for the wake part given in (\ref{eq:wake_couette}), (\ref{eq:wake_channel}) and (\ref{eq:wake_pipe}) for Couette, Channel and Pipe flow respectively. It immediately follows, under the limit $Re_\tau \rightarrow \infty$ or $R^+ \rightarrow \infty$, the wake component vanishes in (\ref{eq:wake1}) and the fractional order is truly universal for any given geometry. This is shown numerically by subtracting the wake contribution in the fractional order, indeed from fig.~\ref{fig:alp_wake} we see a universal behaviour.

\begin{equation} \label{eq:wake1}
    \alpha ~=~ \alpha_{universal} ~+~ \alpha_{wake}
\end{equation}

\begin{equation} \label{eq:wake_couette}
    ^{couette} \alpha_{wake} = 0.08052~(y^+)^{-0.075}~exp[-(y^+/Re_\tau)^{-1}]
\end{equation} 

\begin{equation} \label{eq:wake_channel}
    ^{channel} \alpha_{wake} = 0.36461~(y^+)^{-0.165}~exp[-(y^+/Re_\tau)^{-1.5}]
\end{equation} \

\begin{equation} \label{eq:wake_pipe}
    ^{pipe} \alpha_{wake} = 0.3838~(y^+)^{-0.125}~exp[-(y^+/Re_\tau)^{-1.25}]
\end{equation}

\begin{figure}[H]
\centering
    \subfloat[Couette]{{\includegraphics[trim=0.3cm 0.3cm 0.3cm 0.3cm,clip, width=0.33\textwidth]{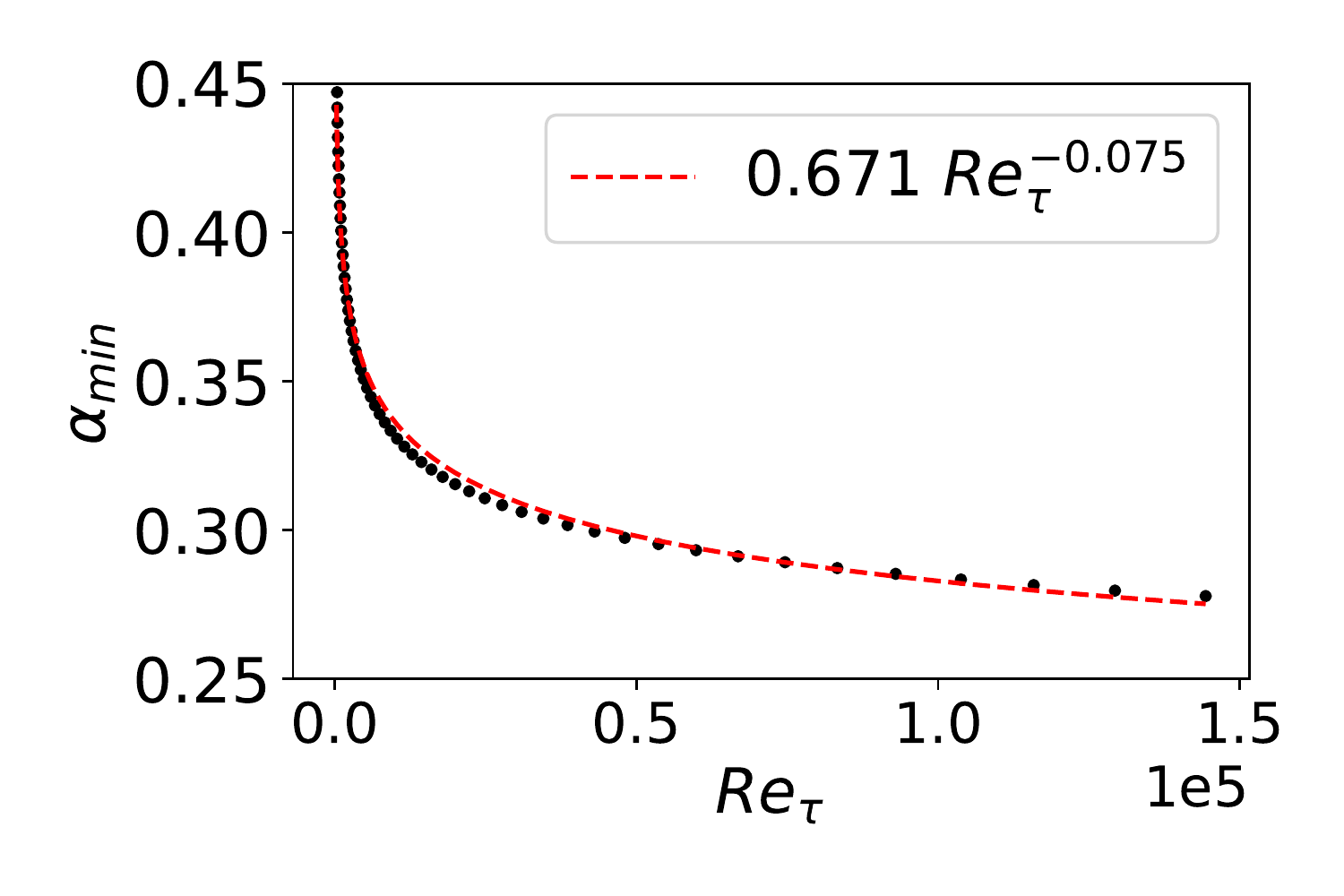}}}%
   \subfloat[Channel]{{\includegraphics[trim=0.3cm 0.3cm 0.3cm 0.3cm,clip,    width=0.33\textwidth]{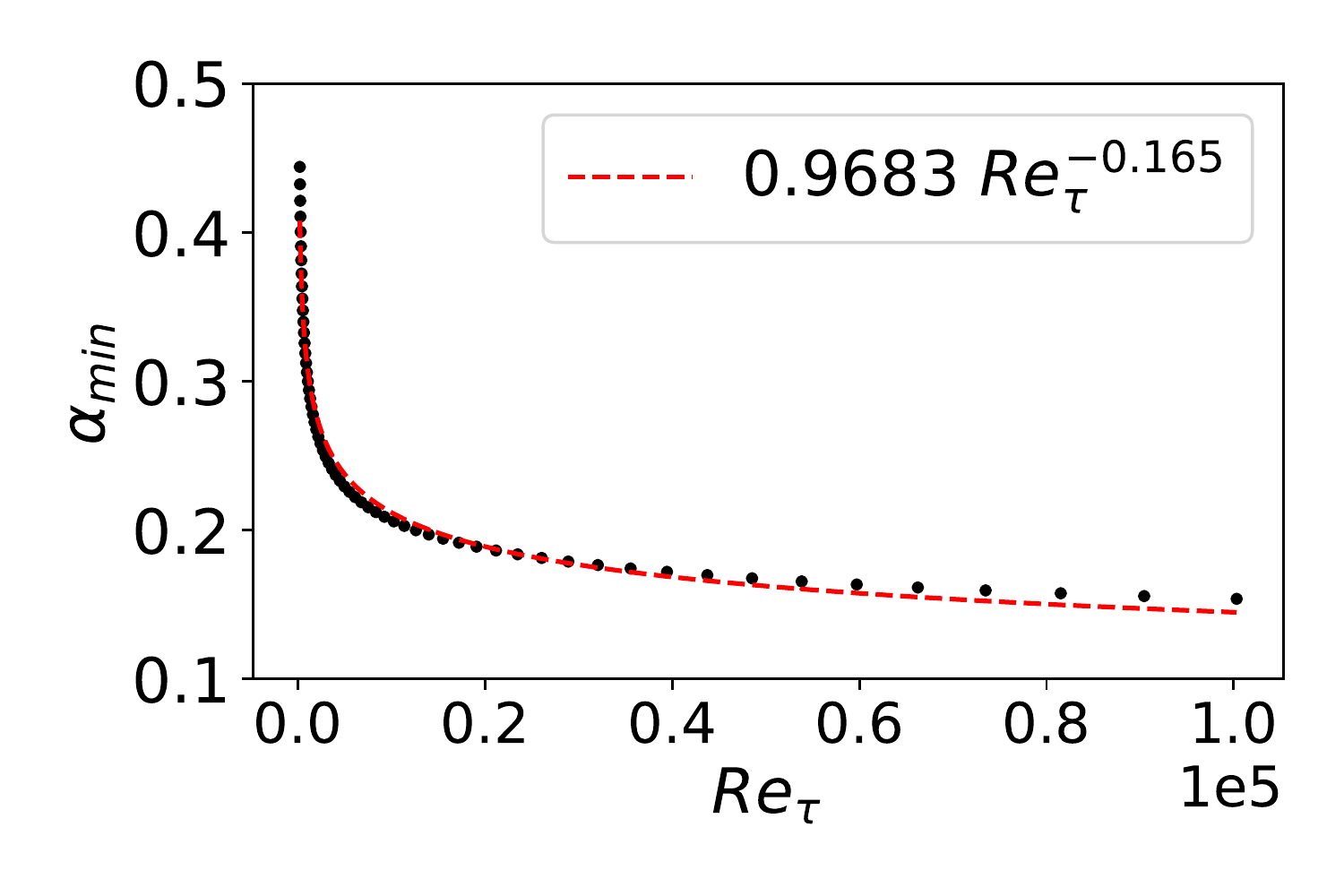}}}%
    \subfloat[Pipe]{{\includegraphics[trim=0.3cm 0.3cm 0.3cm 0.3cm,clip,    width=0.33\textwidth]{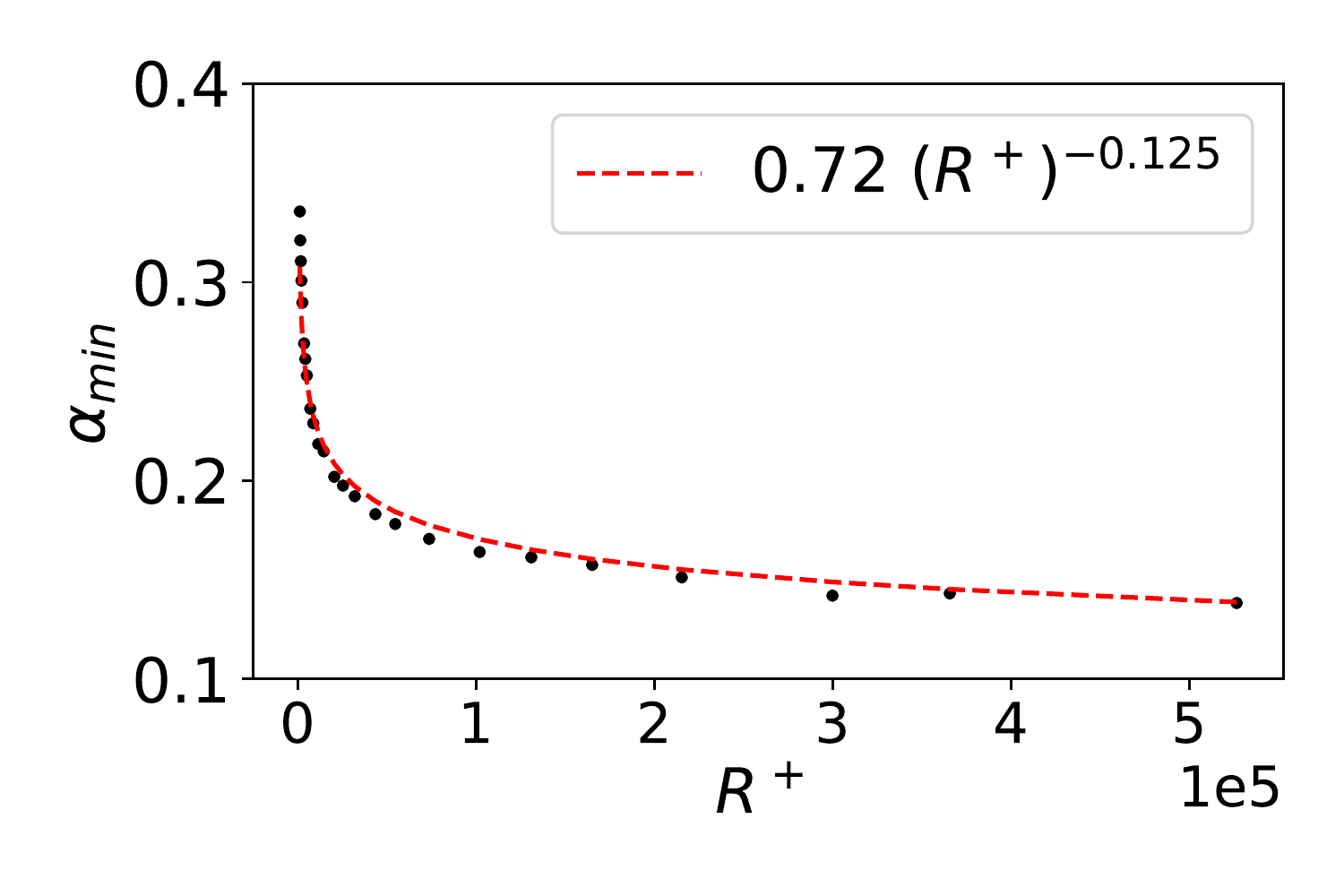}}}%
    \caption{Power Law fitting for Channel, Couette and Pipe shown in red-dashed line for minimum fractional order of two-sided model for respective friction Reynolds number / Karman number shown as black dots}
    \label{fig:alp_asym}
\end{figure}

Furthermore, the fractional order asymptotes for high Reynolds number for all three cases. The initial proposition of Chen \cite{chen2006speculative} of constant fractional order can be deduced from our model, since it asymptotes at high Reynolds number, also the viscous stress is negligible and the only contribution to total shear stress is from the Reynolds Stresses. Since the curve is expected to asymptote at high Reynolds number, the computation of $\alpha_{min}$ from the above discussions is sufficient to prove the same. $\alpha_{min}$ occurs at the centreline; however, since Channel and Pipe has an anomaly at the centreline, these values where computed at $y^+/Re_\tau \approx 0.75$ and $ (1-r)^+/R^+ \approx 0.75 $, respectively but this should not be a problem since all curves are flat at high Reynolds numbers. We employ the Spalding relationship to generate velocity profiles for the case of Couette and Channel, while the Superpipe data along with Spalding relationship for near wall region is used for the case of Pipe. Fig.~(\ref{fig:alp_asym}) not only shows asymptotic relationship but also a power-law relationship. Finally, we report the $\alpha_{min}$ from our computations, which are, Couette: $Re_\tau \approx 144338, \alpha_{min} \approx 0.28 $, Channel: $Re_\tau \approx 100319, \alpha_{min} \approx 0.15 $ and Pipe: $R^+ \approx 526360, \alpha_{min} \approx 0.13$.

\begin{figure}[H]
\centering
    \subfloat{{\includegraphics[trim=0.3cm 0.3cm 0.3cm 0.3cm,clip, width=0.5\textwidth]{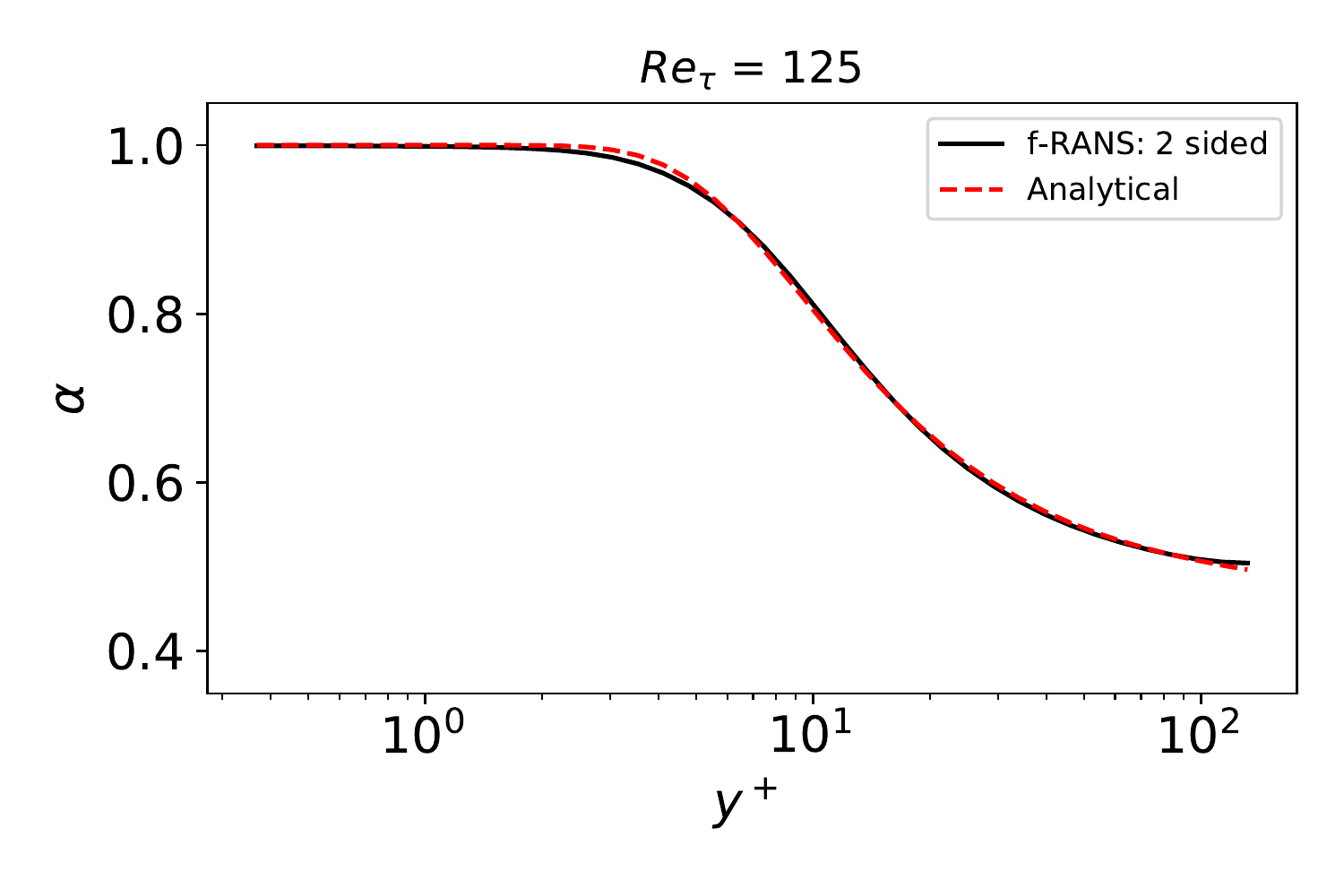}}}%
   \subfloat{{\includegraphics[trim=0.3cm 0.3cm 0.3cm 0.3cm,clip,    width=0.5\textwidth]{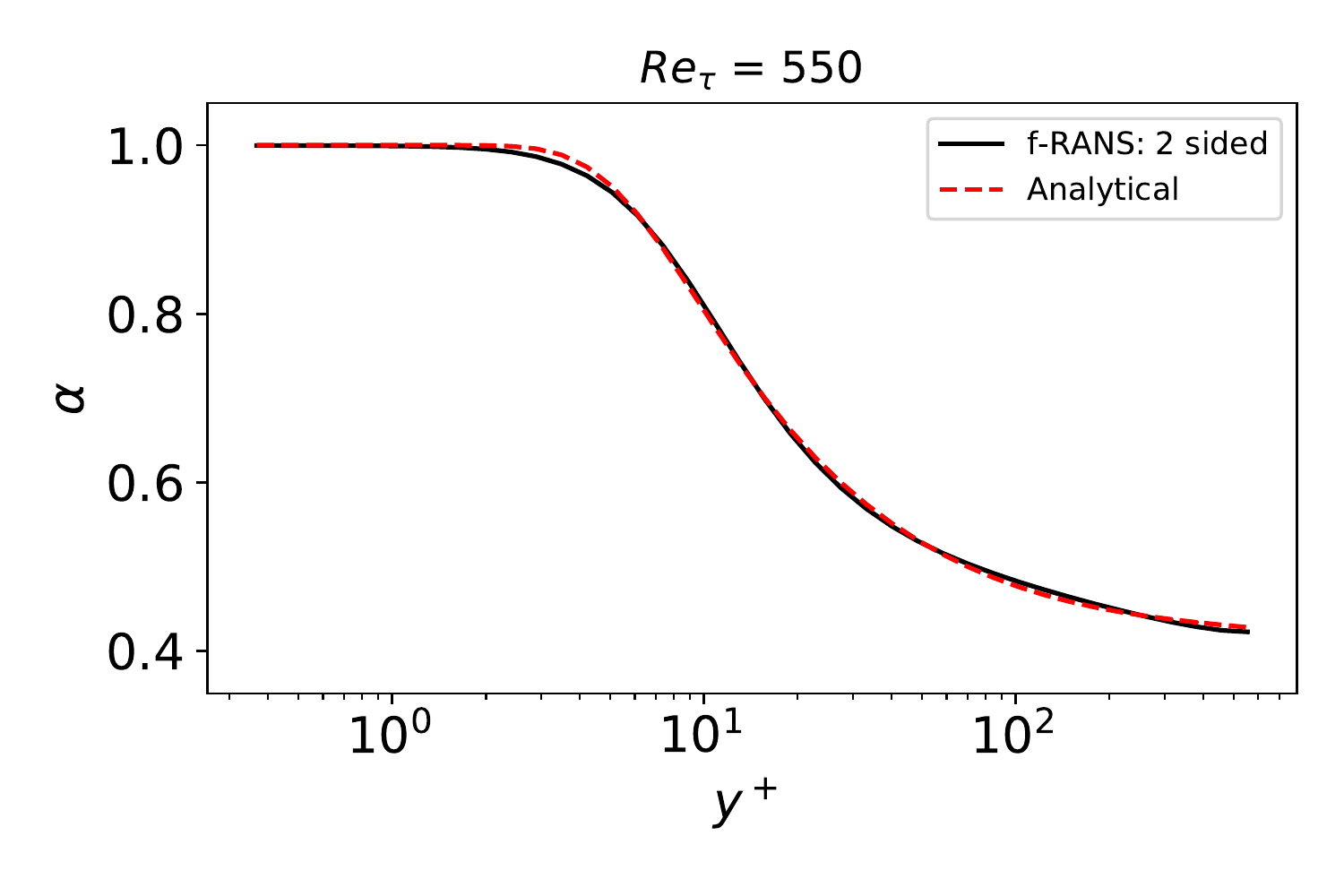}}}%
    \caption{Comparisons of analytical expression for fractional order of two-sided model obtained via fitting to  fractional order computed via inverse modeling for Couette flow}
    \label{fig:couette_two_ydel_anal}
\end{figure}

\begin{figure}[H]
\centering
    \subfloat{{\includegraphics[trim=0.3cm 0.3cm 0.3cm 0.3cm,clip, width=0.5\textwidth]{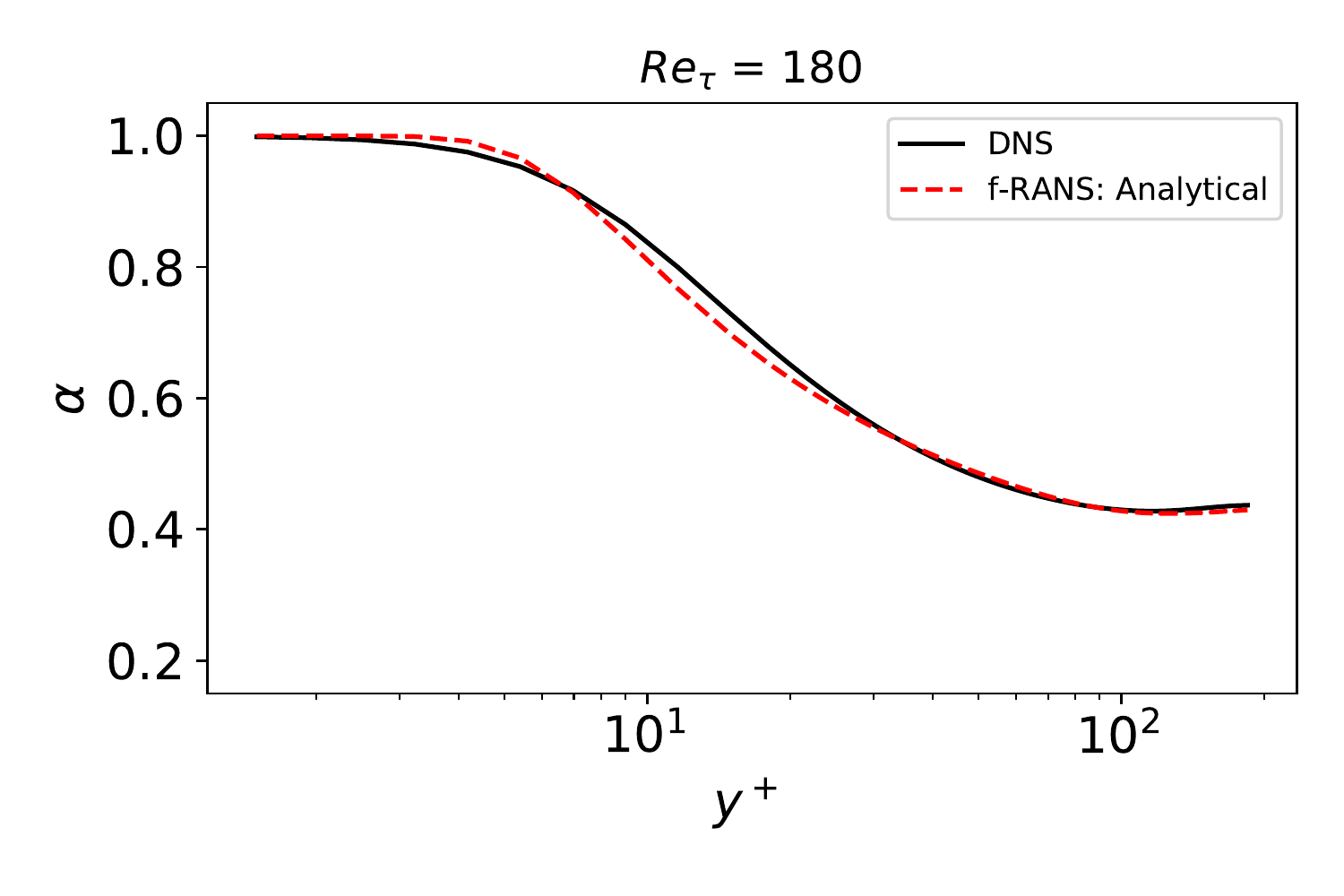}}}%
   \subfloat{{\includegraphics[trim=0.3cm 0.3cm 0.3cm 0.3cm,clip,    width=0.5\textwidth]{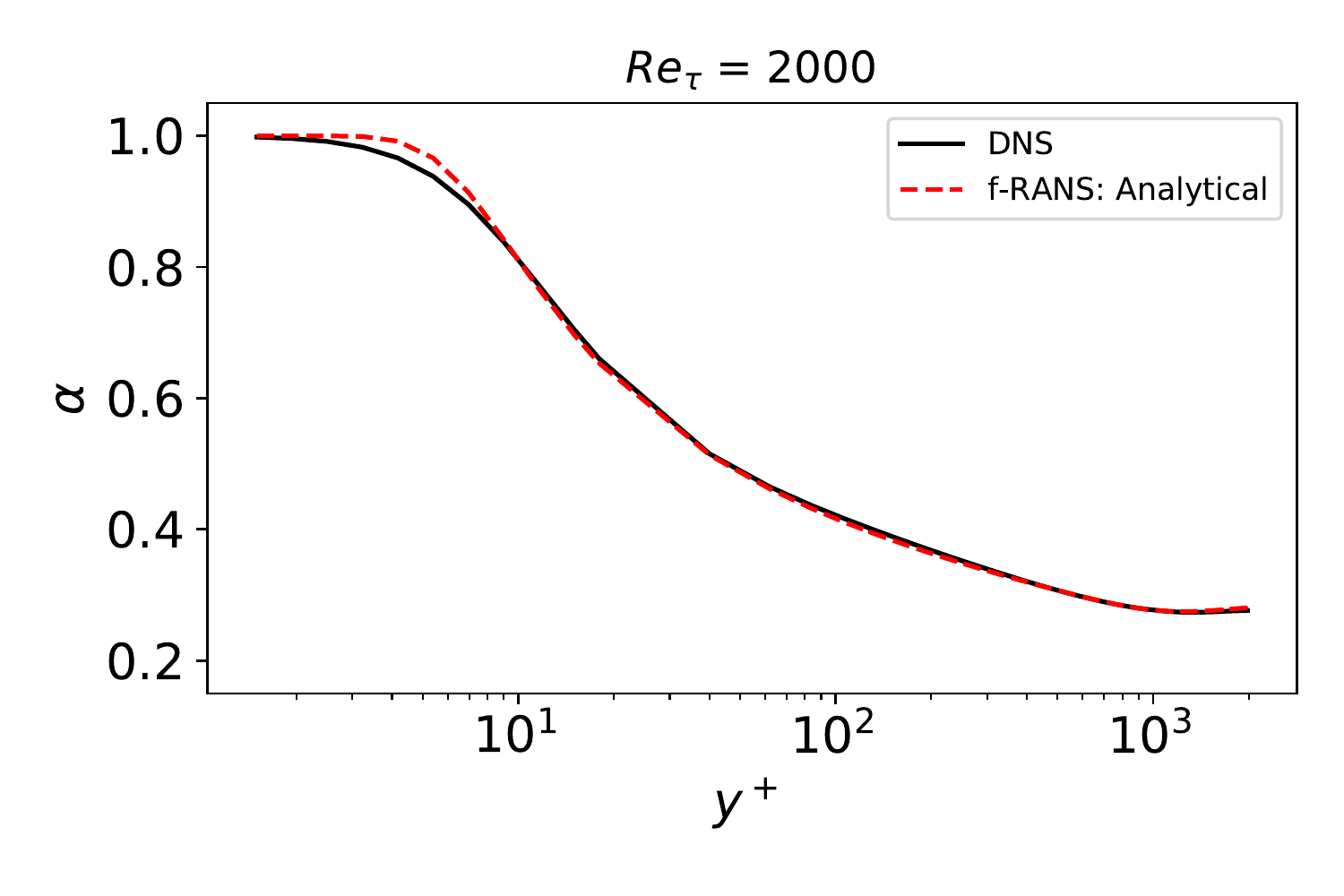}}}%
    \caption{Comparisons of analytical expression for fractional order of two-sided model obtained via fitting to  fractional order computed via inverse modeling for Channel flow }
    \label{fig:channel_two_ydel_anal}
\end{figure}

\begin{figure}[H]
\centering
    \subfloat{{\includegraphics[trim=0.3cm 0.3cm 0.3cm 0.3cm,clip, width=0.5\textwidth]{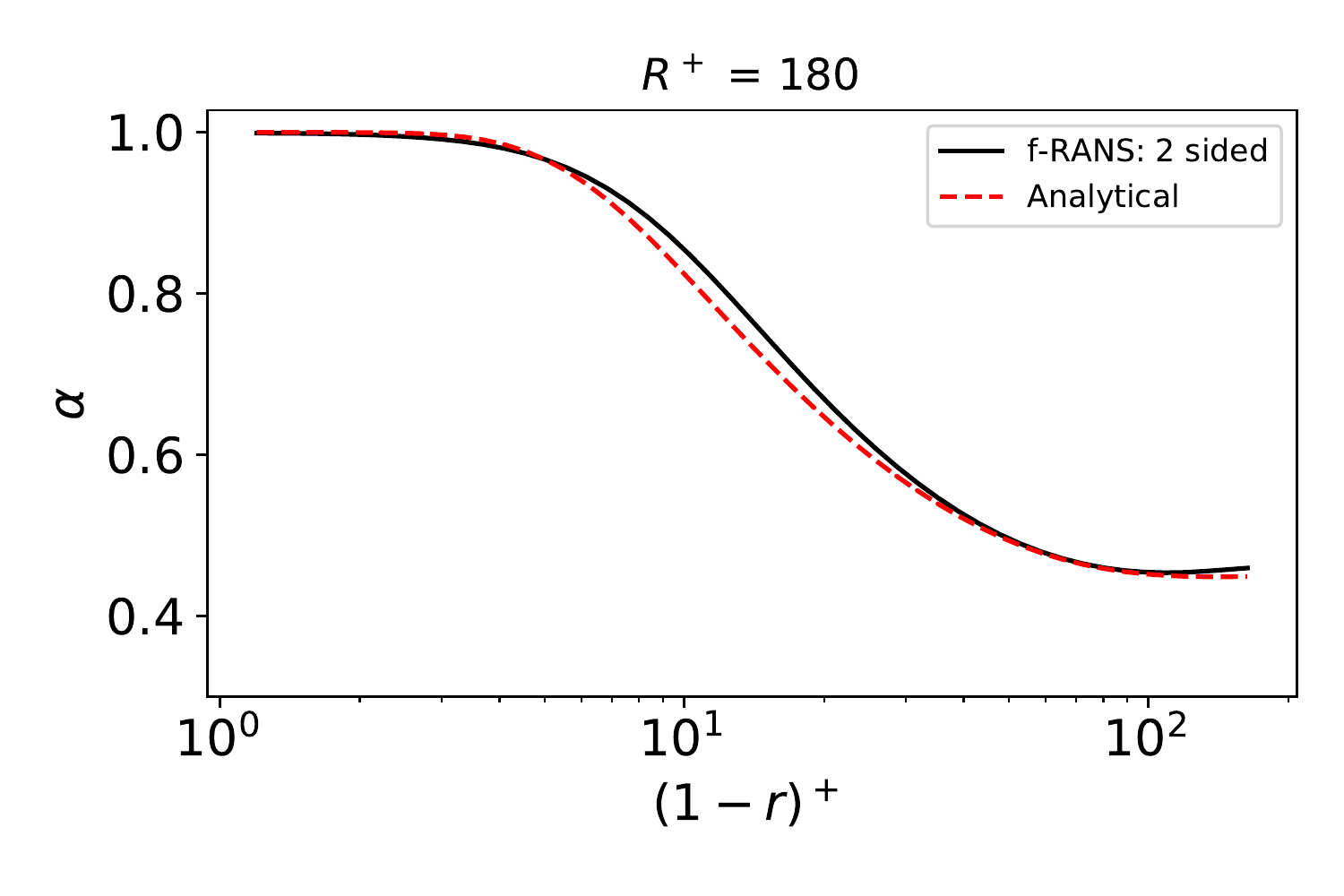}}}%
   \subfloat{{\includegraphics[trim=0.3cm 0.3cm 0.3cm 0.3cm,clip,    width=0.5\textwidth]{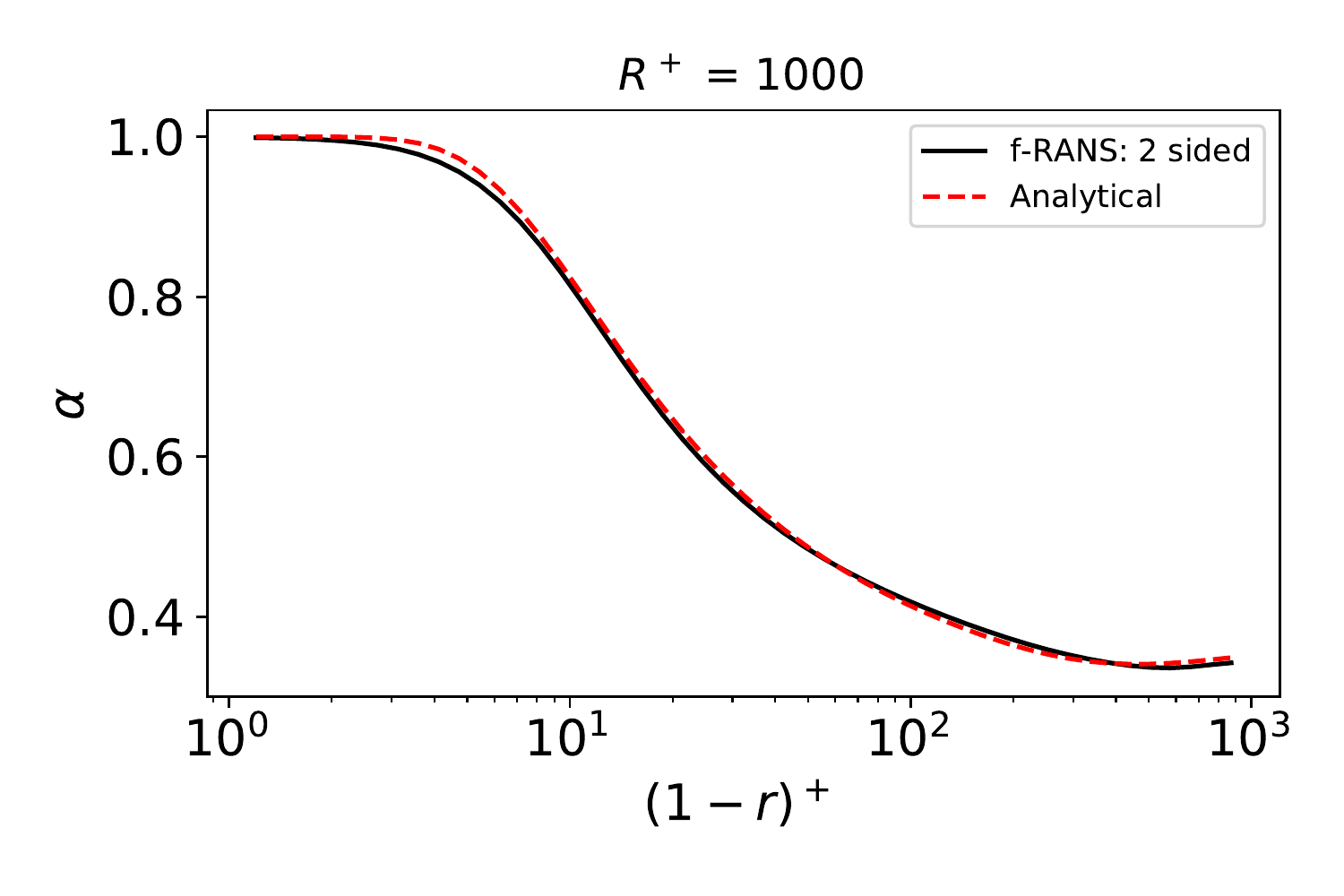}}}%
    \caption{Comparisons of analytical expression for fractional order of two-sided model obtained via fitting to  fractional order computed via inverse modeling for Pipe flow }
    \label{fig:pipe_two_ydel_anal}
\end{figure}

\begin{figure}[H]
\centering
    \subfloat[Couette]{{\includegraphics[trim=0.3cm 0.3cm 0.3cm 0.3cm,clip, width=0.33\textwidth]{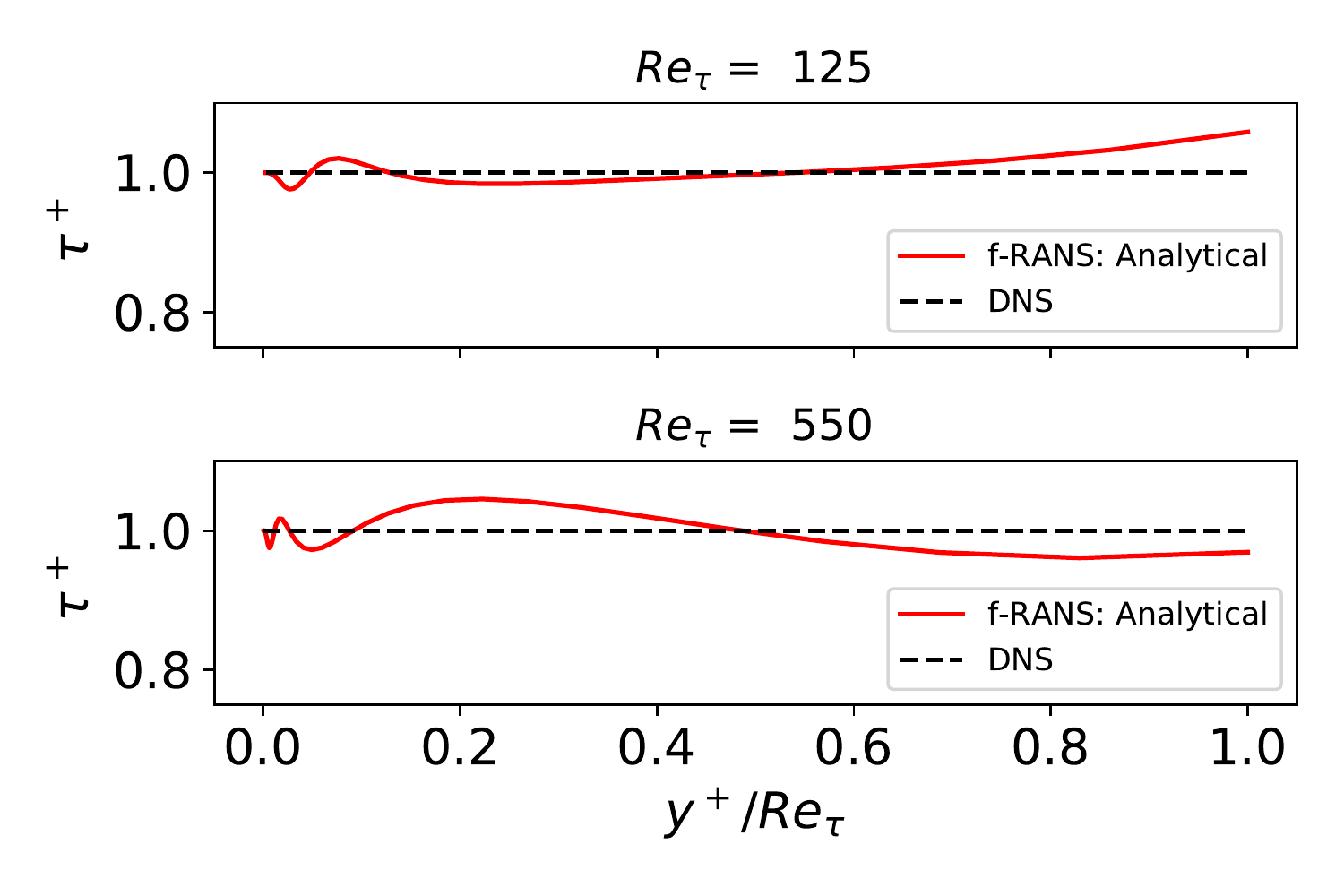}}}%
   \subfloat[Channel]{{\includegraphics[trim=0.3cm 0.3cm 0.3cm 0.3cm,clip,    width=0.33\textwidth]{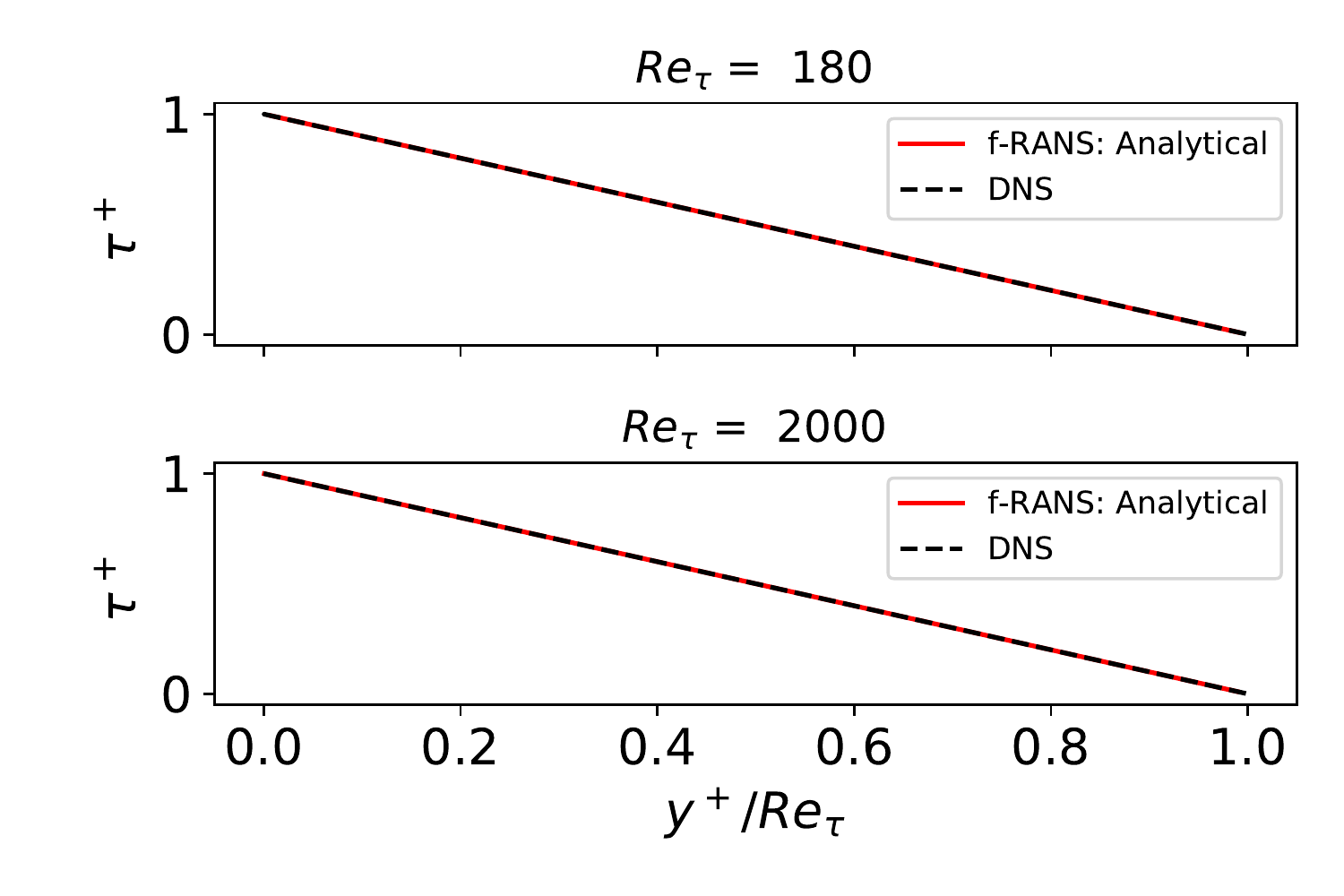}}}%
    \subfloat[Pipe]{{\includegraphics[trim=0.3cm 0.3cm 0.3cm 0.3cm,clip,    width=0.33\textwidth]{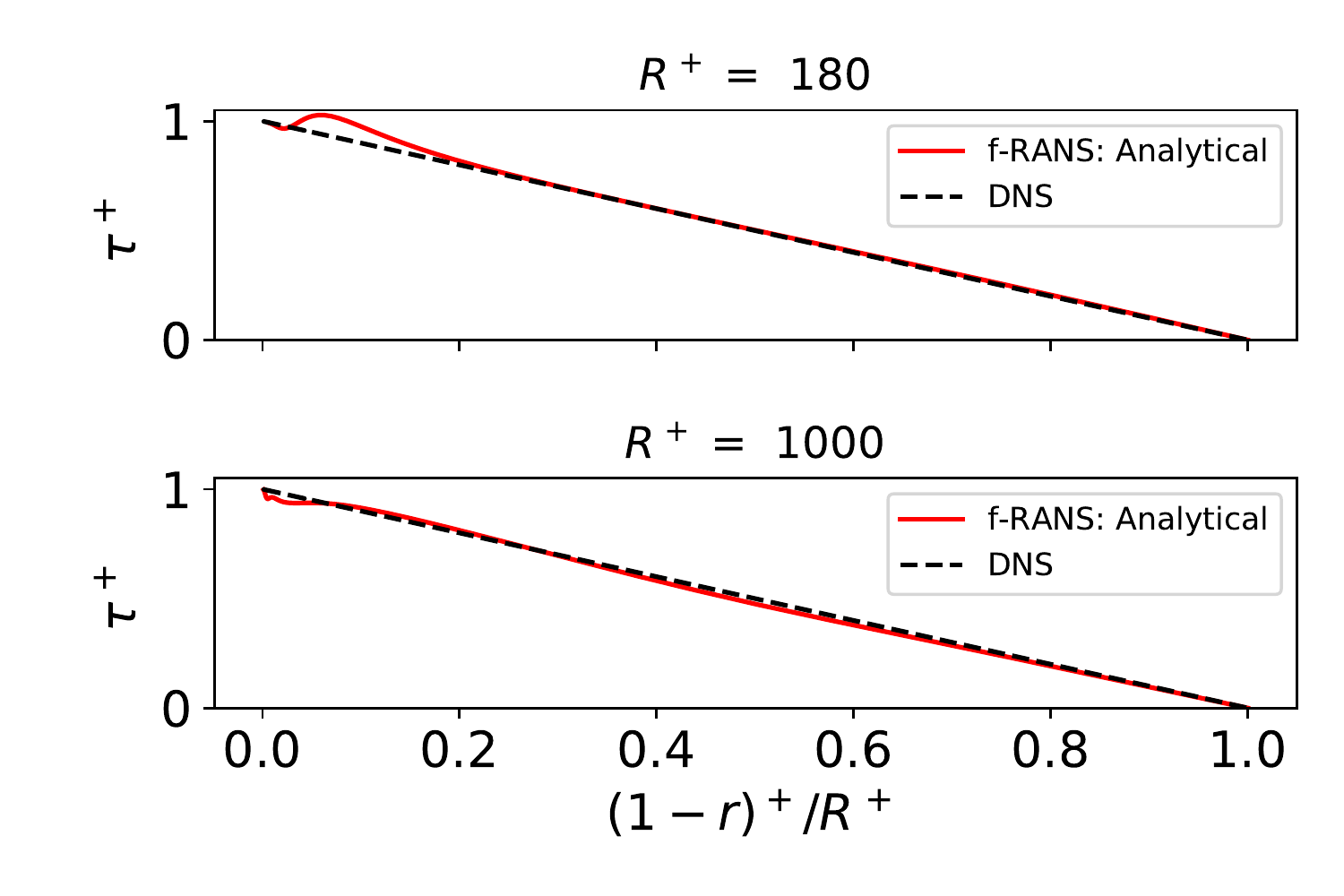}}}%
    \caption{Comparison of total Shear Stress for two-sided f-RANS model using analytical expression for fractional order with DNS databases for Couette, Channel and Pipe flow}
    \label{fig:tau_cc_ana}
\end{figure}

\begin{figure}[H]
\centering
    \subfloat[Couette]{{\includegraphics[trim=0.3cm 0.3cm 0.3cm 0.3cm,clip, width=0.33\textwidth]{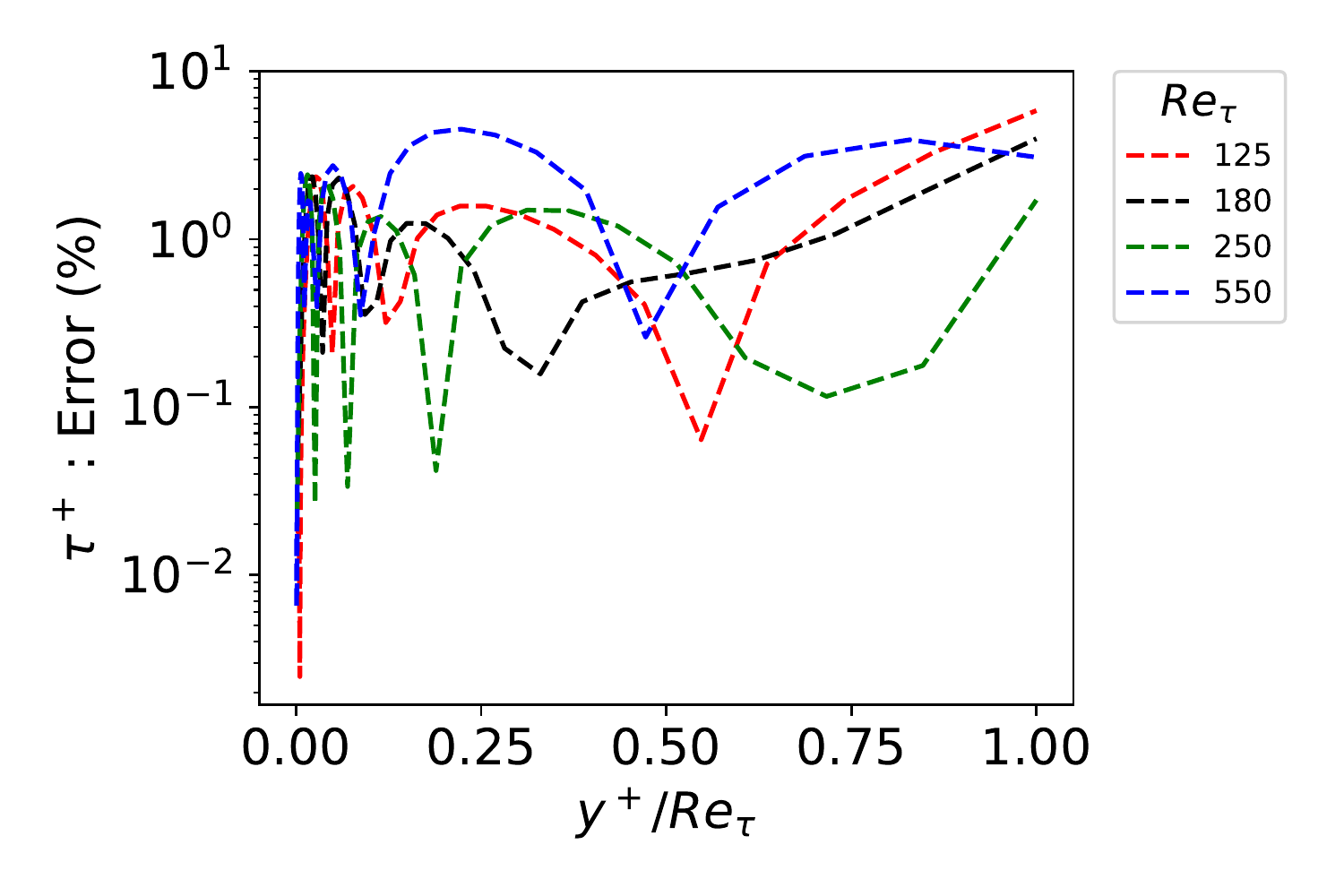}}}%
   \subfloat[Channel]{{\includegraphics[trim=0.3cm 0.3cm 0.3cm 0.3cm,clip,    width=0.33\textwidth]{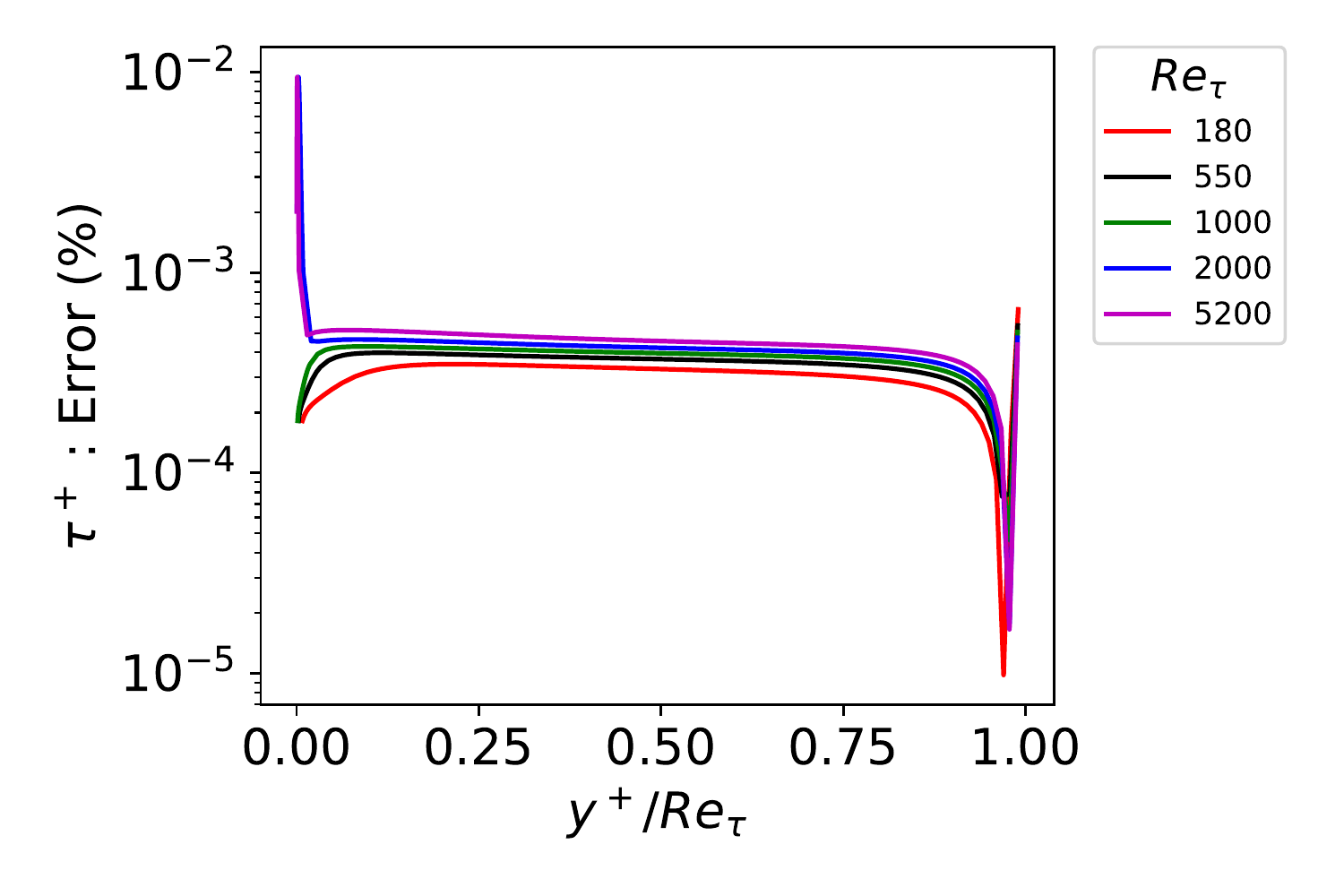}}}%
   \subfloat[Pipe]{{\includegraphics[trim=0.3cm 0.3cm 0.3cm 0.3cm,clip,    width=0.33\textwidth]{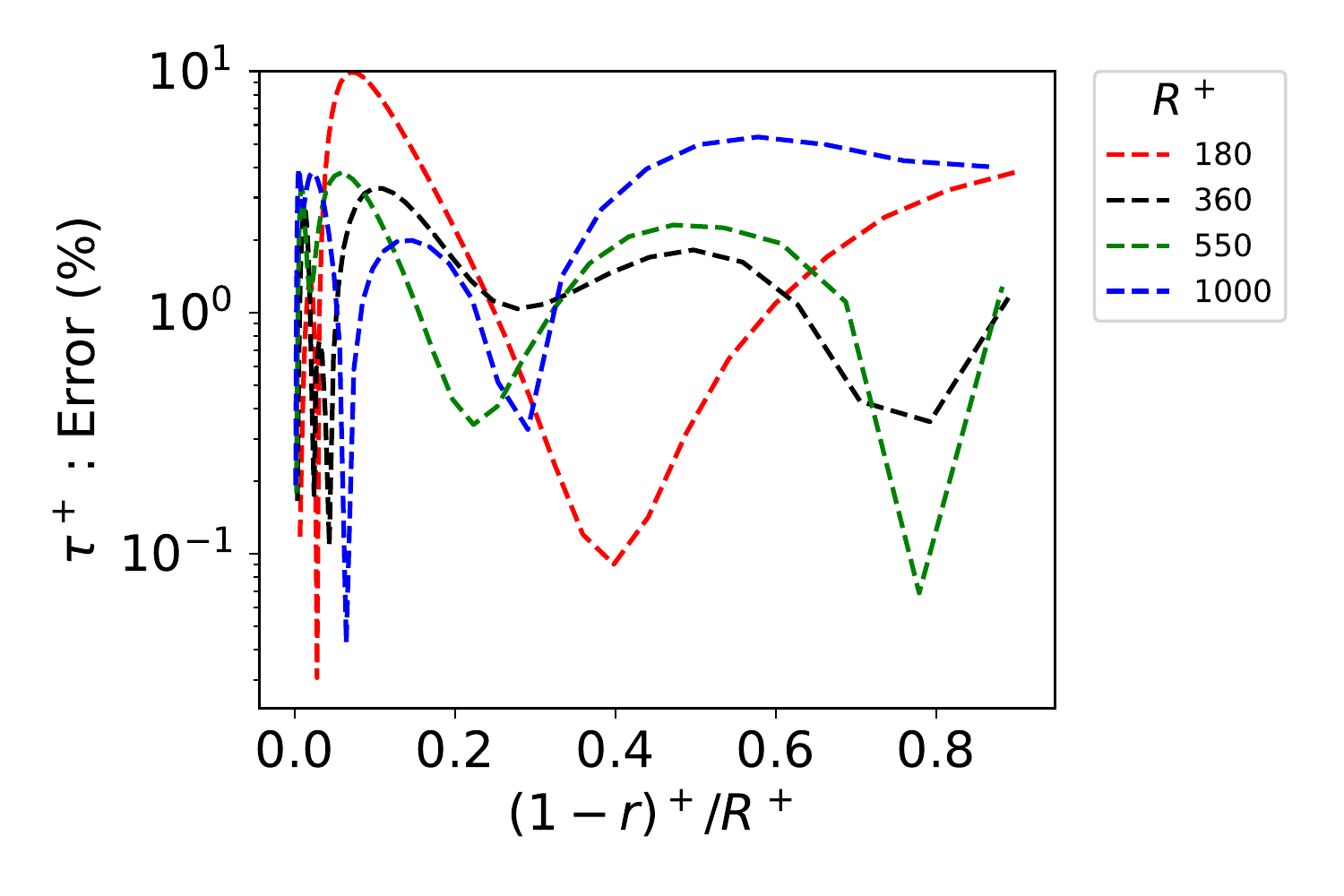}}}%
    \caption{Error of total Shear Stress for two-sided f-RANS model using analytical expression for fractional order with DNS databases for Couette, Channel and Pipe flow}
    \label{fig:tau_cc_err_ana}
\end{figure}

We further fit the fractional order computed for the two-sided model in the preceding section, thus the analytical expression for Couette, Channel and Pipe flow are given in (\ref{eq:anal_couette}), (\ref{eq:anal_channel}) and (\ref{eq:anal_pipe}) respectively. Clearly, there are terms which are depended only on $y^+$ or $r^+$, while the last term is depended on $y^+ / Re_\tau$ or $(1-r)^+ / R^+ $ consistent with our above discussions. Fig.~\ref{fig:couette_two_ydel_anal}, \ref{fig:channel_two_ydel_anal} and \ref{fig:pipe_two_ydel_anal} shows the comparison of the analytical expression with computed fractional order for two-sided model which matches well. We find the mean error for the analytical expression is about $1 \%$ in total shear stress from fig.~\ref{fig:tau_cc_ana} and \ref{fig:tau_cc_err_ana}

\begin{align} \label{eq:anal_couette}
    \begin{split}
         ^{couette} \alpha_{analytical} ~=&~ tanh((6.9/y^+)^{1.116}) ~~+~~ 0.644~(1- tanh((6.9/y^+)^{1.116}))~(y^+)^{-0.0805} \\&+~~ 0.1646~exp[-(y^+/Re_\tau)^{-0.6694}]~(y^+)^{-0.0805}
    \end{split}
\end{align}

\begin{align} \label{eq:anal_channel}
    \begin{split}
         ^{channel} \alpha_{analytical} ~=&~ tanh((6.907/y^+)^{1.5}) ~~+~~ 0.908~(1- tanh((6.907/y^+)^{1.5}))~(y^+)^{-0.175} \\&+~~ 0.418~exp[-(y^+/Re_\tau)^{-1.634}]~(y^+)^{-0.175}
    \end{split}
\end{align}

\begin{align} \label{eq:anal_pipe}
    \begin{split}
         ^{pipe} \alpha_{analytical} ~=&~ tanh((7.988/y^+)^{1.07}) ~~+~~ 0.645~(1- tanh((7.988/y^+)^{1.07}))~(y^+)^{-0.12} \\&+~~ 0.409~exp[-(y^+/Re_\tau)^{-1}]~(y^+)^{-0.12}
    \end{split}
\end{align}

\section{Conclusion} \label{conclusion}
%
%

It is evident that turbulence exhibits a duality of local and non-local regimes, as a result of this duality it has been a challenge to model it. A variable-order fractional model indeed addressed this duality, where the fractional order is unity abiding to local operators for solely viscous-dominated regimes of the flow. Furthermore, with considerations of this duality a unified model form is proposed for all regimes of the flow, namely, laminar, transitional and turbulent. A two-sided model addresses the non-locality in a physical manner by incorporating the aggregate effects from all directions, however, purely on a modeling front, a one-sided model could be advantageous as the computational requirement is less, also for development of a wall model where one may only consider the near wall effects, irrespective of the mean flow. It is to be noted that modeling in wall units, we do not need any additional coefficient to model on ad-hoc basis. As a future work, temporal non-locality or memory effect is to be studied as opposed to the current work of only spatial non-locality. 

\section*{Acknowledgement}

The author dedicates this article in the loving memory of his late father, Shri Pranjivan Ramji Mehta (09 Feb. 1947 -  14 Oct. 2020). This  research  was  funded  by  DARPA (HR00111990025). Research was carried out using computational resources and services at the Center for Computation and Visualization, Brown University. The author is grateful to Prof. George Karniadakis, Brown University for hosting him at Brown University for pursing the current research.

\newpage

\section*{Appendix}

\newpage

\newpage

\newpage

\newpage

\newpage

\bibliographystyle{unsrt}
\bibliography{fractional, turb}

\end{document}